\documentclass[11pt,twoside,a4paper]{article}

\newcommand{\mathsym}[1]{{}}

\newcommand{\baz}{\begin{array}{cc}}
\newcommand{\bad}{\begin{array}{ccc}}
\newcommand{\bi}{\begin{itemize}}
\newcommand{\ei}{\end{itemize}}
\newcommand{\ba}{\begin{array}{c}}
\newcommand{\ea}{\end{array}}

\newcommand{\beqa}{\begin{eqnarray}} 
\newcommand{\eeqa}{\end{eqnarray}} 

\usepackage[utf8]{inputenc}
\usepackage[hang,normal,footnotesize]{caption}
\usepackage{subcaption}
\usepackage{wrapfig}
\usepackage[sort&compress,numbers,colon,merge]{natbib}
\usepackage[usenames,dvipsnames,svgnames,table]{xcolor}

\definecolor{gesfpurple}{rgb}{0.47,0.19,0.42}

\definecolor{gesflanse}{rgb}{0.00,0.50,0.50}

\definecolor{gesfblue}{rgb}{0.08,0.42,0.76}

\definecolor{gesfred}{rgb}{1,0,0}

\definecolor{gesfwhite}{rgb}{1,1,1}

\definecolor{gesfblack}{rgb}{0,0,0}

\newcommand{\gsec}[1]{{\hypersetup{linkcolor=red}Sec.~\ref{#1}\hypersetup{linkcolor=blue}}}

\newcommand{\geqn}[1]{\hypersetup{linkcolor=blue}(\ref{#1})\hypersetup{linkcolor=blue}}
\newcommand{\gfig}[1]{{\hypersetup{linkcolor=violet}Fig.~\ref{#1}\hypersetup{linkcolor=blue}}}
\newcommand{\gtab}[1]{{\hypersetup{linkcolor=gesflanse}Tab.~\ref{#1}\hypersetup{linkcolor=blue}}}

\newcommand{\mee}{\langle m_{ee} \rangle}

\newcommand{\meeLRSM}{\langle m_{ee} \rangle^{\rm LR}_{\rm ee}}
\newcommand{\meeNHmax}{\langle m_{ee} \rangle^{\rm NH}_{\rm max}}

\newcommand{\meeIHmin}{\langle m_{ee} \rangle^{\rm IH}_{\rm min}}

\newcommand{\Ge}{^{76}{\rm Ge}}
\newcommand{\Te}{^{130}{\rm Te}}
\newcommand{\Xe}{^{136}{\rm Xe}}

\usepackage{a4wide}
\usepackage{fancyhdr}
\usepackage{graphicx}
\usepackage{url}
\usepackage{amsfonts}
\usepackage{amsmath,amssymb,amsthm}

\usepackage{mathrsfs}

\usepackage{multirow}
\usepackage{color}
\definecolor{orange}{rgb}{1,0.5,0}
\definecolor{darkred}{rgb}{0.5,0,0}
\definecolor{darkgreen}{rgb}{0,0.5,0}
\definecolor{darkblue}{rgb}{0,0,0.5}

\usepackage{hyperref}
\hypersetup{ colorlinks,
linkcolor=darkgreen,
urlcolor=darkred,
citecolor=darkblue }

\usepackage{cancel}
\usepackage{bbm}
\usepackage{ulem}
\usepackage{color}

\usepackage{units}

\def\be{\begin{equation}}
\def\ee{\end{equation}}

\newcommand{\bea}{\begin{equation} \begin{array}{c}}
\newcommand{\eea}{ \end{array} \end{equation}}

\def\gs{\mathrel{
   \rlap{\raise 0.511ex \hbox{$>$}}{\lower 0.511ex \hbox{$\sim$}}}}
\def\ls{\mathrel{
   \rlap{\raise 0.511ex \hbox{$<$}}{\lower 0.511ex \hbox{$\sim$}}}}

\numberwithin{equation}{section}

\begin{document}

\begin{titlepage}


\begin{center}
 {\huge\sffamily\bfseries\mathversion{bold} 
Half-life Expectations for Neutrinoless Double Beta Decay in Standard and Non-Standard Scenarios}
\\[5mm]
{\large
Shao-Feng Ge\footnote{\texttt{gesf02@gmail.com}}$^{(a)}$,~Werner Rodejohann\footnote{\texttt{werner.rodejohann@mpi-hd.mpg.de}}$^{(a)}$,~Kai Zuber\footnote{\texttt{zuber@physik.tu-dresden.de}}$^{(b)}$\mbox{ } \\
{\small\textit{$^{(a)}$Max-Planck-Institut f\"ur Kernphysik, Saupfercheckweg 1, 69117
Heidelberg, Germany
}}\\
{\small\textit{$^{(b)}$Institut f\"ur Kern- und Teilchenphysik, Technische Universit\"at Dresden, 
01069 Dresden, Germany}}}
\end{center}
\vspace*{1.0cm}

\begin{abstract}
\noindent
We investigate the half-life expectations for neutrinoless double beta decay 
by applying statistical distributions of neutrino mixing observables,  
neutrino mass constraints from cosmology and nuclear matrix elements. The analysis is performed 
in the standard scenario of active Majorana neutrino exchange,  when 
light sterile neutrinos are added, and within TeV-scale left-right symmetric frameworks. The latter two cases correspond to a modified phenomenology of double beta decay for a normal and inverted mass ordering, and thus different discovery potential for future experiments. 

\end{abstract}

\end{titlepage}

\setcounter{footnote}{0}

\section{\label{sec:intro}Introduction}
Neutrinoless double beta decay ($0\nu\beta\beta$) is the main method to search for lepton number violation and its observation would have a variety of profound consequences in particle physics and beyond \cite{Rodejohann:2011mu,Deppisch:2012nb,Pas:2015eia}. 
A large number of experiments is running, under construction or in the planing phase \cite{Schwingenheuer:2012zs,DellOro:2016tmg}. The physics case and goals of the projects 
are usually based on the standard approach to the decay, namely the exchange of 3 light massive Majorana neutrinos. 
The half-life of $0\nu\beta\beta$ then depends on the so-called effective Majorana neutrino  mass, which is a function of seven parameters. 
Those are two Majorana phases that realistically cannot be determined anywhere else, two independent 
mixing matrix elements determined by neutrino oscillations, and three neutrino masses. Those are most strongly constrained by neutrino mass measurements, in particular by cosmology \cite{Wong:2011ip,Hannestad:2016fog}. 

Two natural half-life values for neutrinoless double beta decay are motivated by the inverted neutrino mass ordering. 
In case the inverted mass ordering is realized by nature, the half-life of the decay 
is necessarily finite \cite{Pascoli:2002xq}. If an oscillation experiment shows that the 
neutrino mass ordering is inverted, the decay needs to be observed with a certain half-life; if it is not observed with that half-life, 
neutrinos are Dirac particles. Alternatively and more exotic, they could be pseudo-Dirac particles, or the 
active neutrino contribution is canceled by a new physics diagram. Turning the argument around, 
if double beta decay experiments find a value or limit below the expected maximal 
half-life for the inverted ordering, the mass ordering should be normal (the same alternative cases from above can apply). 
It should be stressed that while for small neutrino masses the normal ordering predicts much larger half-lives, neutrino mass can in fact still be around 0.1 eV, leading to sizable half-lives even for this case. However, for all aspects it must be noted that uncertainties in current 
nuclear matrix element calculations \cite{Engel:2016xgb} render a precise prediction of half-lives difficult. 


We investigate in this paper expectations of effective masses and half-lives for neutrinoless double beta decay. 
Towards this we statistically sample probability densities of available neutrino oscillation and cosmology fits\footnote{We do not consider information from direct neutrino mass experiments \cite{Drexlin:2013lha}, which currently do not provide mass limits that are compatible with cosmology. This will change in the future.}, in order
to obtain a distribution of expected effective masses. Adding probability distributions for the nuclear matrix elements, 
which can be obtained from a particular nuclear matrix element calculation (QRPA), where correlated uncertainties are known \cite{Faessler:2008xj,Faessler:2013hz,Lisi:2015yma} 
(and cover the general range of matrix element calculations with other approaches), we then give probability distributions of double beta decay half-lives for the normal or inverted mass ordering. This allows to estimate the discovery potential of future experiments. 
Similar analyses were recently performed also in Refs.\ \cite{Agostini:2017jim,Caldwell:2017mqu,Zhang:2015kaa}. We note here that a predictive  model may be behind the origin of neutrino mass and lepton mixing, and could give a definite prediction for the effective mass, or at least a non-trivial range of values. In the absence of a clear candidate for such a model, a statistical approach used here and in Refs.\ \cite{Agostini:2017jim,Caldwell:2017mqu,Zhang:2015kaa} is useful. 
What we add here to the discussion of half-life expectations, except for the statistical treatment of matrix elements, is that we consider double beta decay mechanisms beyond the standard approach. 
We take into account the addition of light eV-scale sterile neutrinos and a certain class of TeV-scale left-right symmetric 
theories. 
For the former case the effective mass receives a contribution from a new light Majorana neutrino whose mass and mixing is determined by fits to short-baseline anomalies \cite{Gariazzo:2017fdh}. In case of the TeV-scale 
left-right symmetric models that we consider, the relevant modified "effective mass" still depends on mixing matrix elements and neutrino masses, but in a different form. Both cases thus allow again a statistical treatment towards half-life expectations using the results of oscillation and cosmology fits.  The two 
non-standard scenarios have the interesting feature that the phenomenology for the normal and inverted mass ordering can be quite 
different, thus giving quite different discovery potential for future experiments.  

The rest of the paper is organized as follows: In Section \ref{sec:proc} our procedure to obtain half-life expectations is summarized. It is applied in Sections \ref{sec:SI}, \ref{sec:st} and \ref{sec:LR} to the standard approach to double beta decay, 
light sterile neutrinos and left-right symmetric theories, respectively. 
Section \ref{sec:concl} concludes our paper.



\section{\label{sec:proc}Procedure}


Following here the notation of Refs.\ \cite{Faessler:2008xj,Faessler:2013hz,Lisi:2015yma}, 
the half-life for neutrinoless double beta decay of isotope $i$ is given by 
\begin{equation}\label{eq:master}
(T_{1/2}^{-1})_i = G^x_i \left| {\cal M}_i^x \, p_x\right|^2, 
\end{equation}
where the subscript $x$ depends on the mechanism,  
$G^x_i$ is the phase space factor, ${\cal M}^x_i$ the nuclear matrix element (NME) and $p_x$ a particle physics parameter in units of energy, e.g.\ the effective neutrino Majorana mass, see below. All mechanisms for $0\nu\beta\beta$ considered in this paper depend on low energy neutrino parameters and  have identical phase space factors.  \gtab{tab:NME} summarizes the numbers for matrix elements and phase space factors used in what follows. 
Note that $\eta_i \equiv \log_{10} \mathcal M_i$ is the logarithm of the NME and $\gamma_i \equiv - \log_{10} [G_i/({\rm y}^{-1} \, {\rm eV}^{-1})]$ is the logarithm
of the phase space factor, 
in accordance with \cite{Faessler:2008xj,Faessler:2013hz}. Together with 
the effective mass, $\mu \equiv \log_{10}(\langle m_{ee} \rangle / {\rm eV})$, the
half-life $\tau_i \equiv \log_{10} (T_{1/2}/{\rm y})$ can be expressed as 
$  \tau_i
=
  \gamma_i
- 2 \eta_i
- 2 \mu$.
\begin{table}[t!]
\centering
\begin{tabular}{c|c|c|c|ccc}
  \multirow{2}{*}{$i$} & \multirow{2}{*}{$\gamma_i$} & \multirow{2}{*}{$\eta^0_i$} & \multirow{2}{*}{$\sigma_i$} & \multicolumn{3}{c}{correlation matrix $\rho_{ij}$} \\
  & & & & $\Ge$ & $\Te$ & $\Xe$ \\
\hline
  $\Ge$ & 25.517 & 0.635 & 0.122 & 1 \\
  $\Te$ & 24.674 & 0.498 & 0.158 & 0.899 & 1 \\
  $\Xe$ & 24.644 & 0.254 & 0.187 & 0.805 & 0.916 & 1
\end{tabular}
\caption{The log of light neutrino exchange phase space factor $\gamma_i$, the central value of the log of the NME
         $\eta_i$ and its error $\sigma_i$, together with the (symmetric)
         error correlation matrix $\rho_{ij}$ \cite{Faessler:2008xj}.}
\label{tab:NME}
\end{table}
\noindent 
The probability distribution functions (PDF) are sampled by the NuPro package \cite{NuPro} according to the total likelihood $\mathcal L_{\rm tot}$: 
\begin{equation}
  \mathcal L_{\rm tot}
=
  \mathcal L_{\rm osc}
\times
  \mathcal L_{\rm cos}
\times
  \mathcal L_{\rm NME} \,,
\end{equation}
which is a product of the contributions
from neutrino oscillation fits ($\mathcal L_{\rm osc}$), 
cosmological constraints ($\mathcal L_{\rm cos}$), and a statistical determination
of NMEs ($\mathcal L_{\rm NME}$). We do not include information from direct neutrino mass experiments \cite{Drexlin:2013lha}, which are currently much weaker than the cosmology constraint and would not affect our results. 

\begin{figure}[t]
\begin{center}
\includegraphics[width=6cm, height=7.7cm,angle=-90]{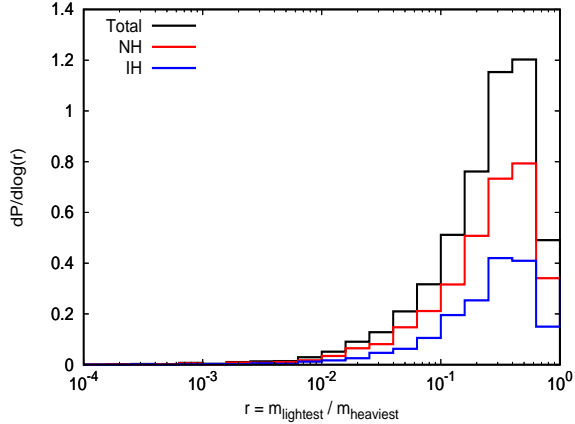} 
\caption{Probability distribution for the ratio of the smallest and largest neutrino mass.}
\label{fig:massratio}
\end{center}
\end{figure}

In what regards the future ranges of the oscillation parameters, we will use except 
noted otherwise the very narrow ranges that JUNO \cite{JUNO} and other experiments will have
determined. 
In addition to constraints on mixing angles and the two
mass-squared differences, the oscillation constraint 
\begin{equation}
  \mathcal L_{\rm osc}
=
  \mathcal L(\theta_{13})
  \mathcal L(\theta_{12})
  \mathcal L(\Delta m^2_a)
  \mathcal L(\Delta m^2_s)
  e^{- \Delta \chi^2_{\rm MH}/2} \,,
\end{equation}
also includes the current preference for the normal ordering with $\Delta \chi^2_{\rm MH} = 3.6$ \cite{Capozzi:2017ipn}. If we consider a contribution from light eV-scale sterile neutrinos to the effective mass we furthermore add to $\mathcal L_{\rm osc}$ 
the distributions of the relevant new mixing matrix element and mass-squared difference, 
$\mathcal L(|U_{e4}|)$ and $\mathcal L(\Delta m^2_{41})$ from Ref.\ \cite{Gariazzo:2017fdh}. 
The cosmological data also shows some preference for the normal ordering. 
Figure 1 of Ref.\ \cite{Hannestad:2016fog} shows the probability distributions of the smallest neutrino mass for both mass orderings, when current cosmology data are fitted (CMB, BAO and local measurements of the Hubble parameter). Integrating over those curves and taking the ratio of the  integrals weighted by the prior probabilities for the orderings (assumed to be equal), gives 
the posterior probabilities for the orderings, namely $ {\rm NH \!: IH}  = 2:1$. 
We add to this discussion in our \gfig{fig:massratio} the distribution of the ratio of the smallest mass 
and the largest neutrino mass. Ratios below $10^{-2}$ are not likely. 

We use the decomposed PDF curves from Figure 1 of \cite{Hannestad:2016fog} to first sample the neutrino mass ordering and then the smallest mass $m_{1,3}$. The first step can be realized by sampling a random number $0 \leq r \leq 1$ and determine the mass ordering  according to 
\begin{equation}
  r
\lessgtr
  \frac {P_{\rm NH}} {P_{\rm NH} + e^{- \Delta \chi^2_{\rm MH} / 2} P_{\rm IH}} \,,
\label{eq:PMH}
\end{equation}
for normal and inverted, respectively; 
$\Delta \chi^2_{\rm MH} = 3.6$ represents the preference from neutrino oscillation
measurements \cite{Capozzi:2017ipn}, corresponding to odds of $ {\rm NH \!: IH} = 1 : e^{-\Delta \chi^2_{\rm MH}/2}  \simeq 6:1$. The  cosmology limit of $\sum_i m_i <  0.14\,\mbox{eV}$ 
at 95\% CL has posterior odds of $ {\rm NH \! : IH} = 2:1$ \cite{Hannestad:2016fog}, so in total the odds in favor of NH are about $12:1$. 
After sampling the neutrino mass hierarchy
according to \geqn{eq:PMH} as the first step, the neutrino mass scale
is then sampled according to one of the two PDF curves,
$\mathcal L(m_1)$ for NH or $\mathcal L(m_3)$ for IH, respectively.
The effective mass also depends on the Majorana phases, which are sampled linearly between 0 and $2\pi$.


%
Using the numbers from  \gtab{tab:NME} we can sample also the NMEs. First we sample 3 independent Gaussian distributions in the diagonalized basis of $\sigma_i \rho_{ij} \sigma_j$ and then rotate the 3 sampled eigenvalues
back to the real basis. To be precise, the likelihood of the NMEs is defined as 
\begin{equation}
  \mathcal L_{\rm NME}
\equiv
  \exp \left[ - \frac 1 2 \sum_{ij} (\eta_i - \eta^0_i) \Sigma^{-1}_{ij} (\eta_j - \eta^0_j) \right] ,
\quad \mbox{with} \quad
  \Sigma_{ij}
\equiv
  \sigma_i \rho_{ij} \sigma_j \,.
\end{equation}
The correlated
error matrix $\Sigma$ is symmetric and can be diagonalized as
$\Sigma^{-1} = U^T \, \overline \Sigma^{-1} U$, where $\overline \Sigma^{-1}$ is diagonal while
$U$ is a unitary mixing matrix. Correspondingly, the eigenbasis
$\overline \eta = U \eta$ is a linear combination of the original variables.
In the eigenbasis, the original likelihood $\mathcal L_{\rm NME}$ can be
decomposed into separate Gaussian distributions
\begin{equation}
  \mathcal L_{\rm NME}
=
  \exp \left[ - \frac 1 2 \sum_i (\overline \eta_i - \overline \eta^0_i) \overline \Sigma^{-1}_{ii} (\overline \eta_i - \overline \eta^0_i) \right]
=
  \Pi_i \exp \left[ - \frac 1 2 (\overline \eta_i - \overline \eta^0_i) \overline \Sigma^{-1}_{ii} (\overline \eta_i - \overline \eta^0_i) \right] .
\end{equation}
After sampling the separate Gaussian distributions of $\overline \eta$,
the value of the original NMEs can be obtained by transforming to the original
basis, $\eta = U^{-1} \, \overline \eta$. 
When sampling the half-life in \gsec{sec:LR}, where both light and heavy neutrinos (masses much larger than the neutrino 
momentum of about 100 MeV in double beta decay) can contribute, we use for the sake of 
consistency the NMEs from \cite{Lisi:2015yma} as summarized in
\gtab{tab:NME-LRSM}. 
\begin{table}[t]
\centering
\begin{tabular}{c|ccc}
 $\eta$ & ${}^{76}{\rm Ge}$ & ${}^{130}{\rm Te}$ & ${}^{136}{\rm Xe}$ \\
\hline
 light & $0.6 \pm 0.32$ & $0.504 \pm 0.32$ & $0.267 \pm 0.32$ \\
 heavy & $2.4 \pm 0.25$ & $2.364 \pm 0.25$ & $2.135 \pm 0.25$
\end{tabular}
\caption{The logarithms of the NMEs for light and heavy neutrinos \cite{Lisi:2015yma}.}
\label{tab:NME-LRSM}
\end{table}

We focus in this paper on the isotopes $^{76}$Ge, $^{130}$Te and $^{136}$Xe, which are subject to strong future limits 
\cite{fut} by LEGEND \cite{legend}, SNO+ \cite{Prior:2017nuf} and CUORE/CUPID \cite{Wang:2015raa,Wang:2015taa}, as well as nEXO \cite{Mong:2016sza} and KamLAND-Zen \cite{Brunner:2017iql}, respectively. 
While there are of course many NME calculations available \cite{Engel:2016xgb}, we stick in what follows to the ones from 
Refs.\ \cite{Faessler:2008xj,Lisi:2015yma} as given in \gtab{tab:NME} and \gtab{tab:NME-LRSM}. However, the implied range of values covers most of the matrix element values from different calculations \cite{Engel:2016xgb}. 
Moreover, the distributions of effective masses are of course not affected by this, while the half-life distributions are subject to small changes, while their shapes are very robust. Furthermore, we do not take into account the possibility of "quenching" \cite{Barea:2013bz}, i.e.\ the potential effective 
reduction of the axial coupling constant $g_A$ for neutrinoless double beta decay from its bare value 1.27. It would imply that the half-lives, which depend approximately on the fourth power of $1/g_A$, would increase.

\section{\label{sec:SI}Neutrinoless Double Beta Decay in the Standard Approach}

\subsection{\label{sec:nu}Relation to Neutrino Physics}
\begin{figure}[t!]
\centering
\includegraphics[height=0.45\textwidth,width=0.3529\textwidth,angle=-90]{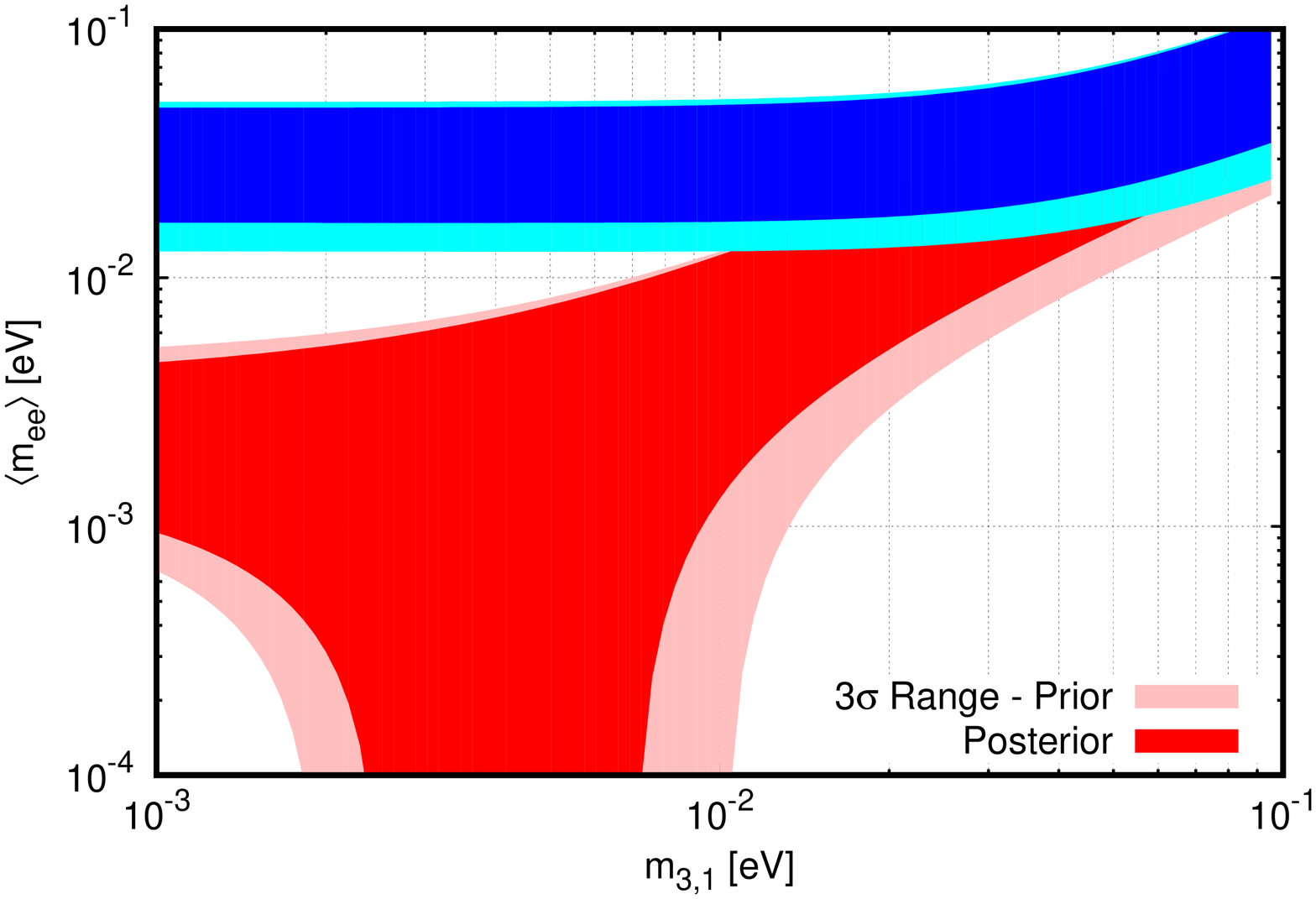}
\includegraphics[height=0.45\textwidth,width=0.3529\textwidth,angle=-90]{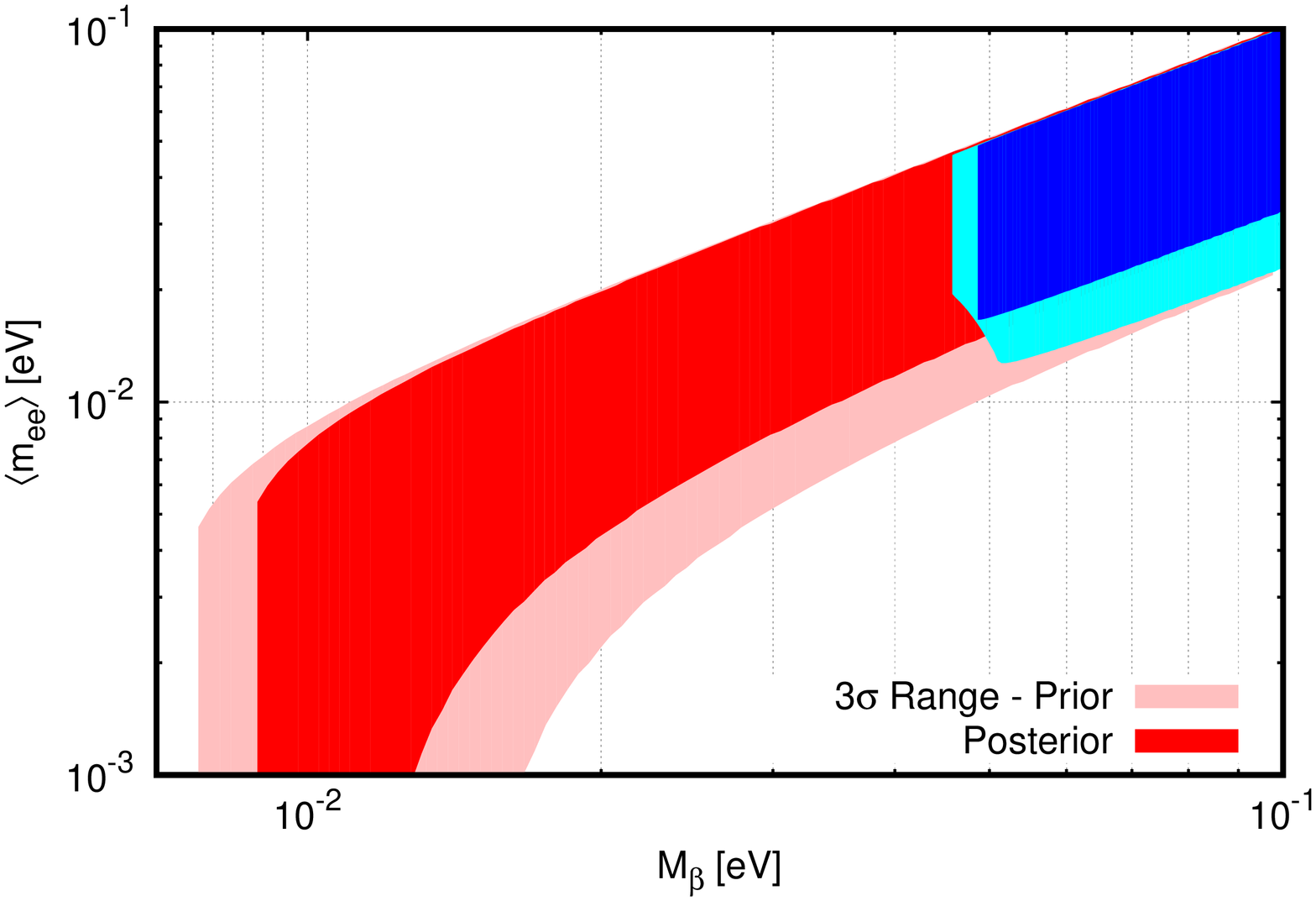}
\includegraphics[height=0.45\textwidth,width=0.3529\textwidth,angle=-90]{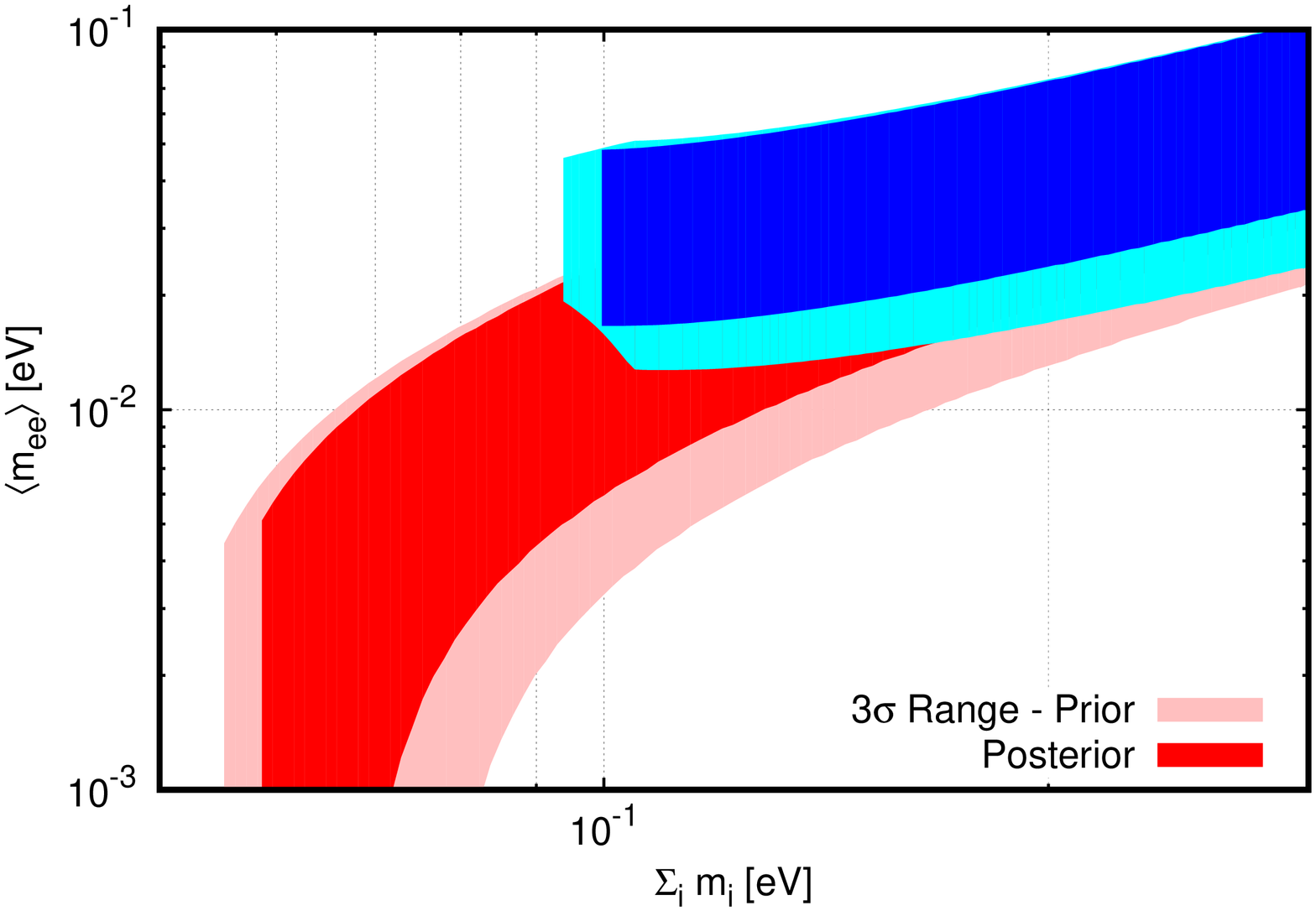}
\caption{The effective mass as a function of the smallest neutrino mass ($m_1$ or $m_3$), $\beta$--decay mass $M_\beta=\sqrt{|U_{ei}|^2 m_i^2}$, and the sum of mass eigenvalues, within the $3\sigma$ range before 
         (prior) and after (posterior) JUNO. The inverted ordering (blue) always predicts a non-vanishing value of the effective mass.}
\label{fig:obs}
\end{figure}

The decay width of neutrinoless double beta decay in the standard approach 
is proportional to the square of the effective Majorana neutrino mass (in short, effective mass)
\begin{equation}
\langle m_{ee} \rangle = \left| U_{e1}^2 \,m_1 + U_{e2}^2 \,m_2 \,e^{i \alpha} + U_{e3}^2 \,m_3 \,e^{i \beta} \right|, 
\end{equation}
where $U_{ei}$ are elements of the PMNS matrix, $m_i$ the neutrino mass eigenstates determined by the smallest mass ($m_1$ for the normal, $m_3$ for the inverted ordering),  $\Delta m^2_{\rm a}$ $(\Delta m^2_{\rm s})$ the atmospheric (solar) mass-squared 
differences and $\alpha,\beta$ two unknown Majorana phases. We show in \gfig{fig:obs} the effective mass versus the smallest mass and versus the other neutrino mass observables 
$M_\beta = \sqrt{|U_{ei}|^2 \, m_i^2}$ from direct mass experiments \cite{Drexlin:2013lha} and the sum of neutrino masses from cosmology. We use the $3\sigma$ ranges of the oscillation parameters as of now \cite{Capozzi:2017ipn}, as well as with future precision, in particular from JUNO \cite{JUNO}. 

\begin{figure}[ht!]
\centering
\includegraphics[height=0.325\textwidth,width=5.54332cm,angle=-90]{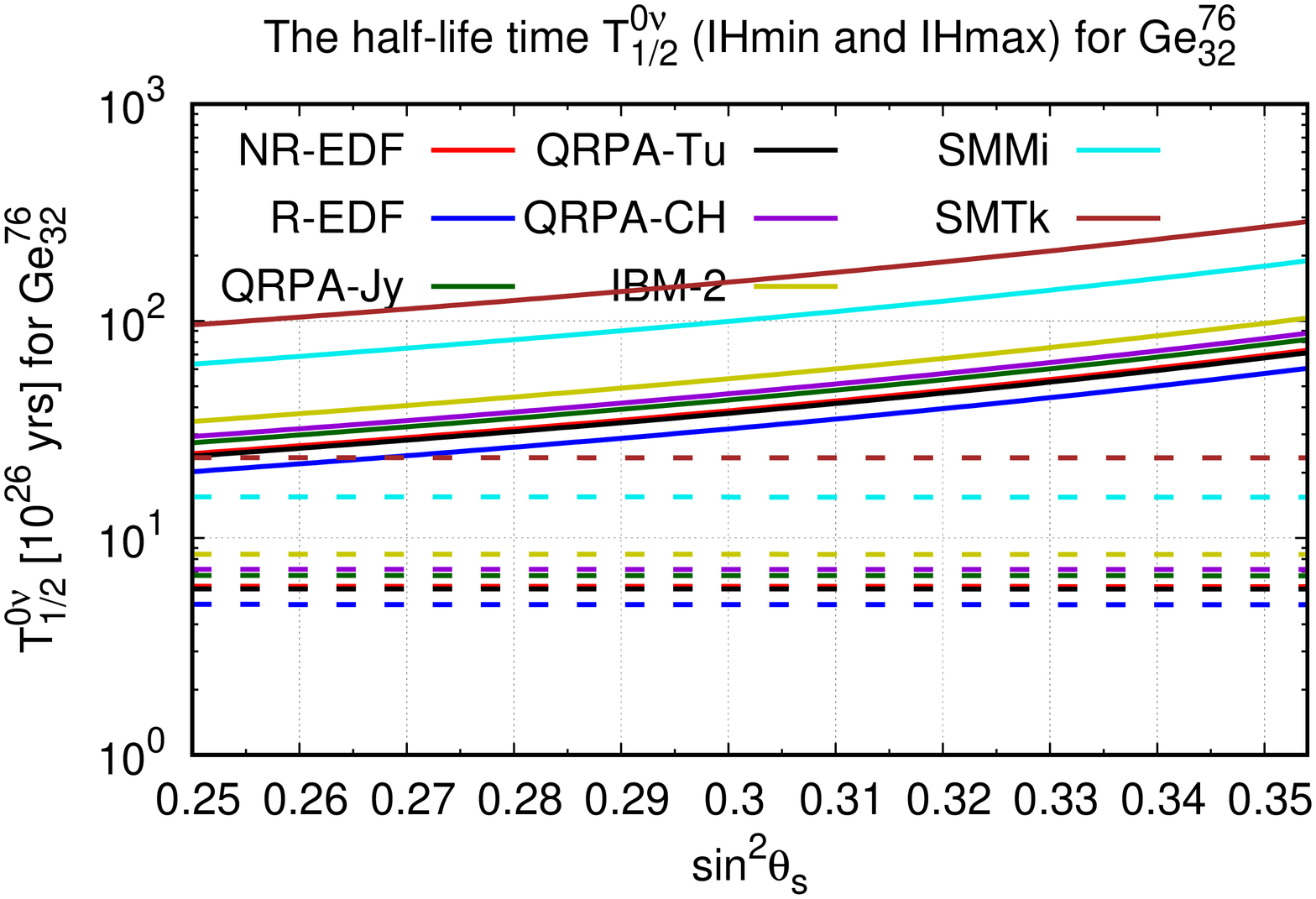}
\includegraphics[height=0.325\textwidth,width=5.54332cm,angle=-90]{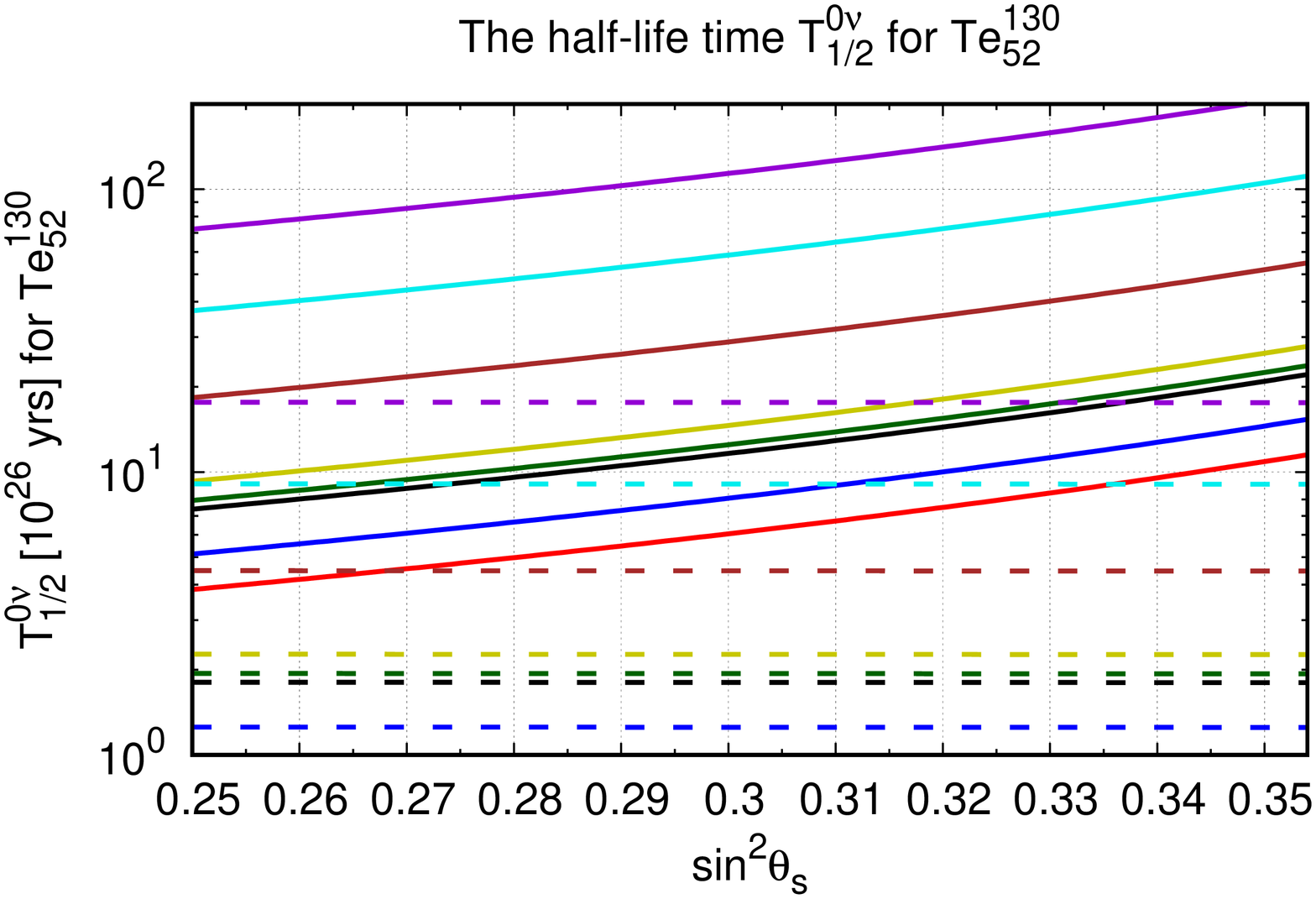}
\includegraphics[height=0.325\textwidth,width=5.54332cm,angle=-90]{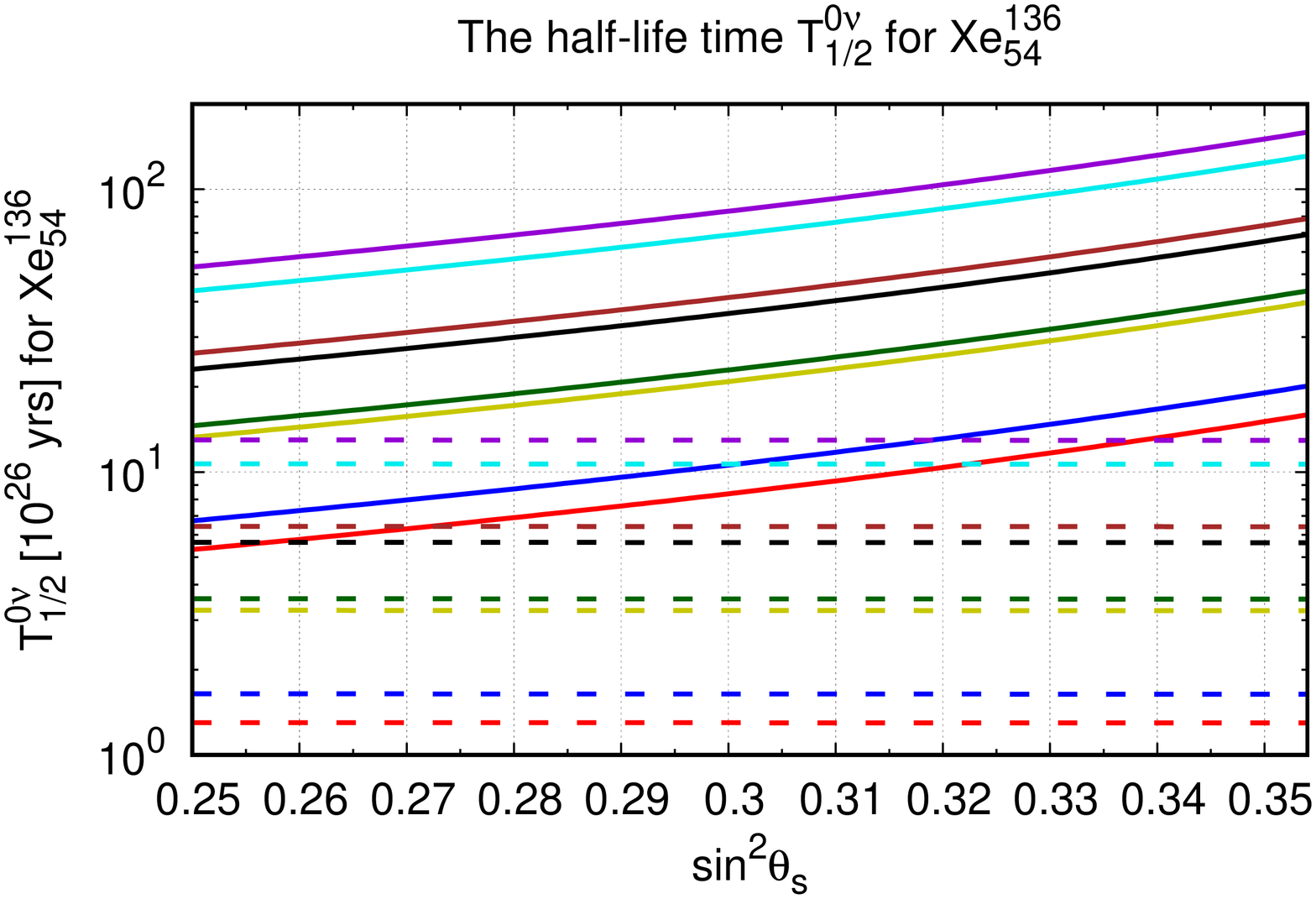}
\includegraphics[height=0.32\textwidth,width=5.54332cm,angle=-90]{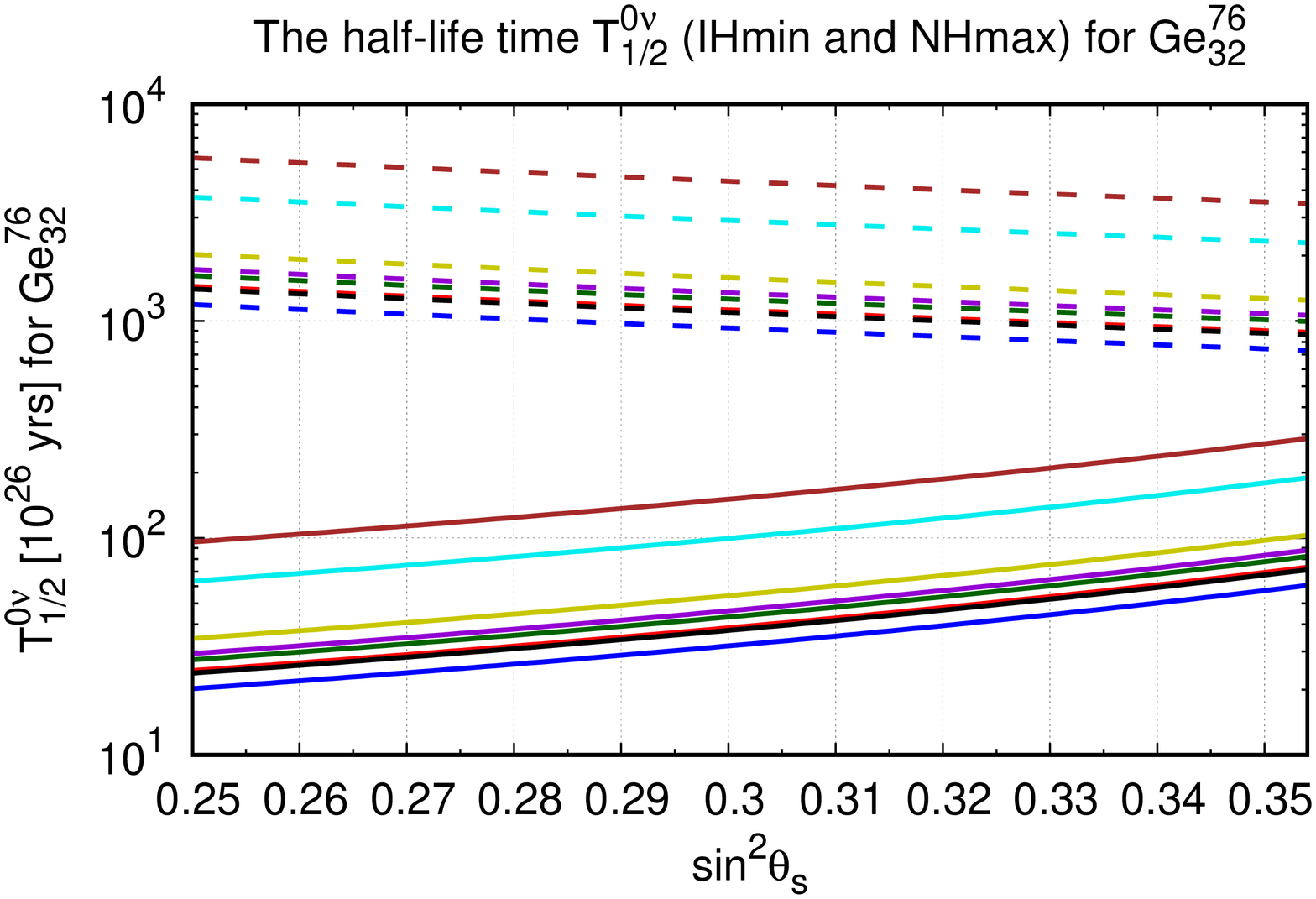}
\includegraphics[height=0.32\textwidth,width=5.54332cm,angle=-90]{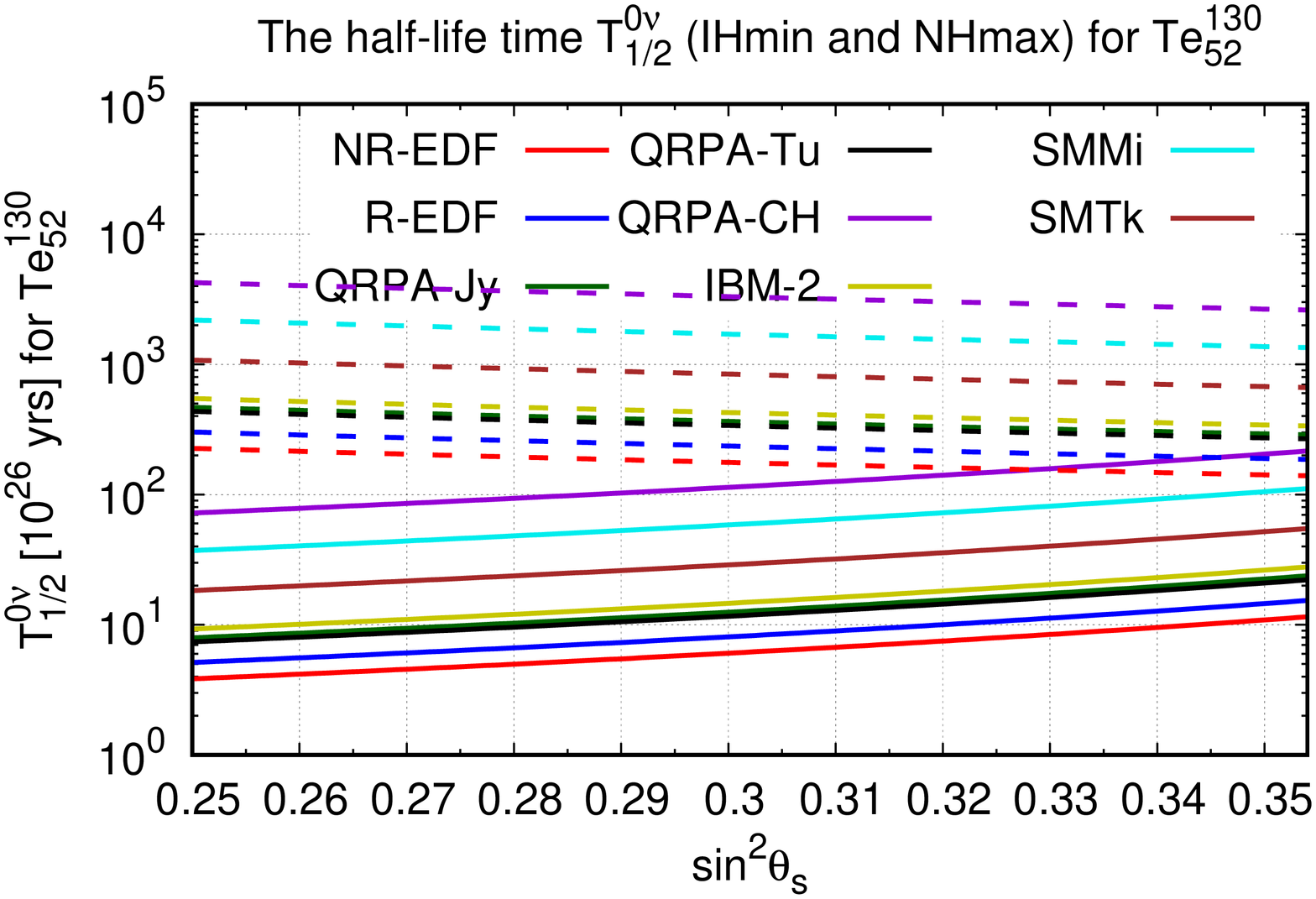}
\includegraphics[height=0.32\textwidth,width=5.54332cm,angle=-90]{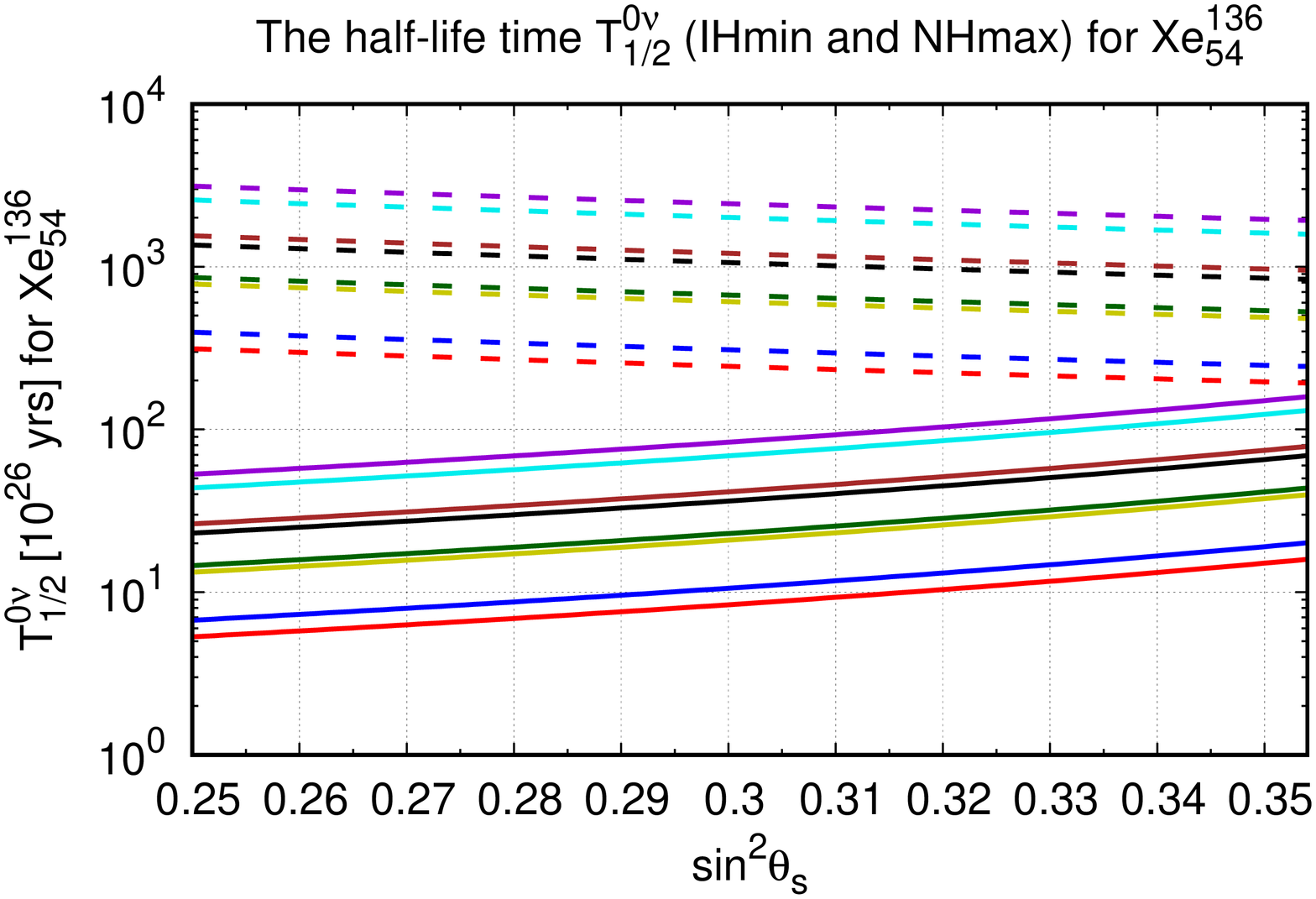}
\includegraphics[height=0.32\textwidth,width=5.54332cm,angle=-90]{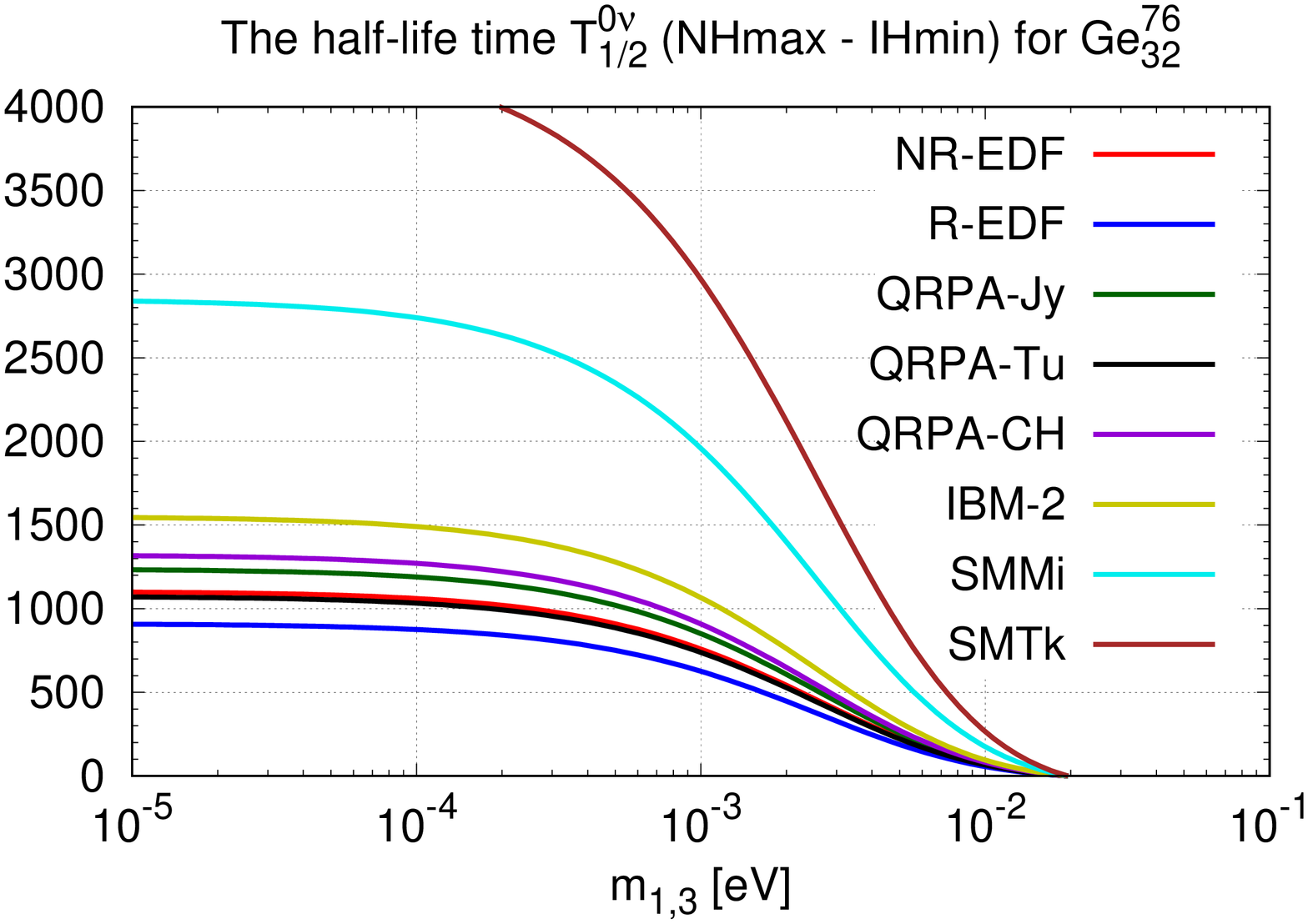}
\includegraphics[height=0.32\textwidth,width=5.54332cm,angle=-90]{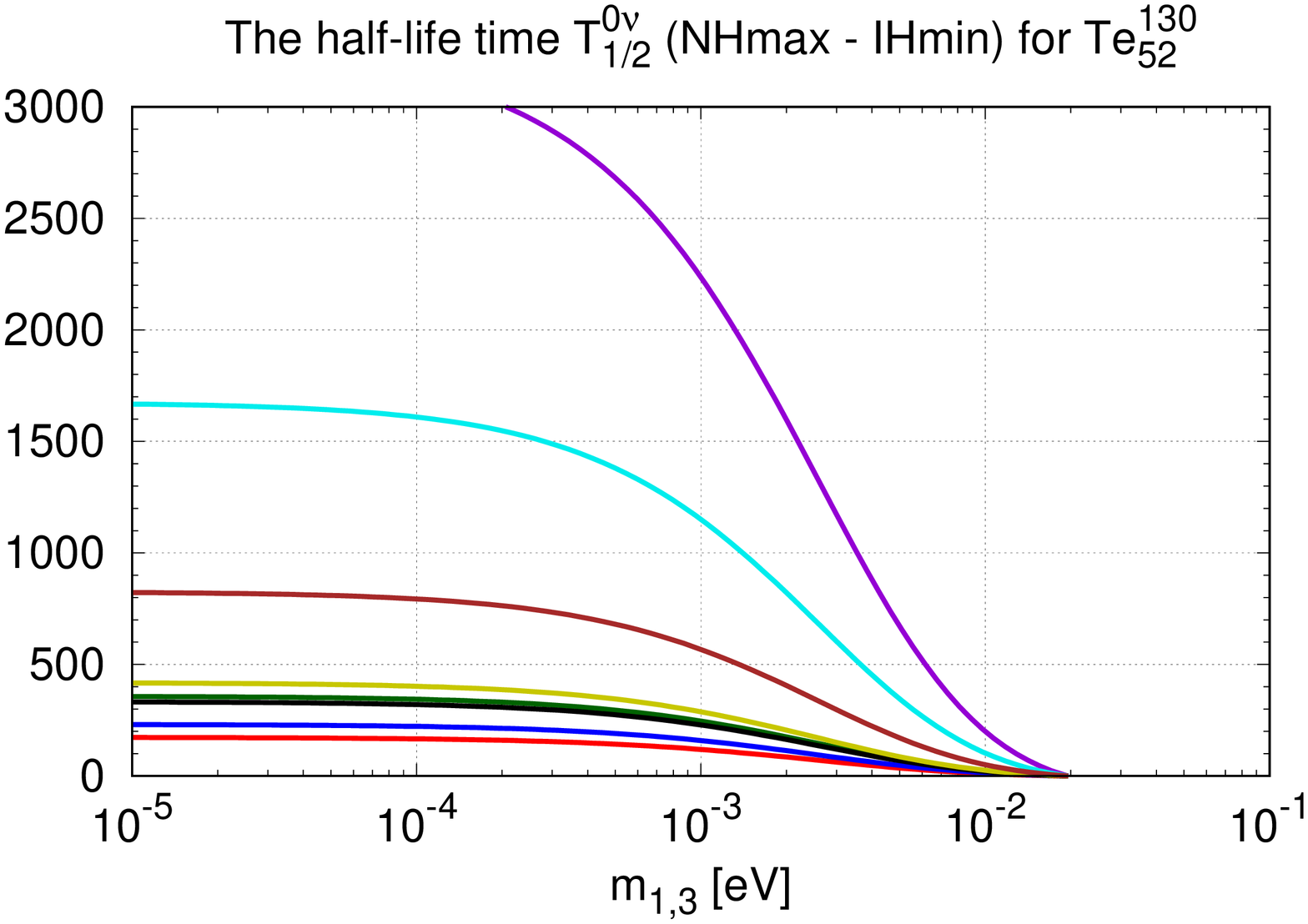}
\includegraphics[height=0.32\textwidth,width=5.54332cm,angle=-90]{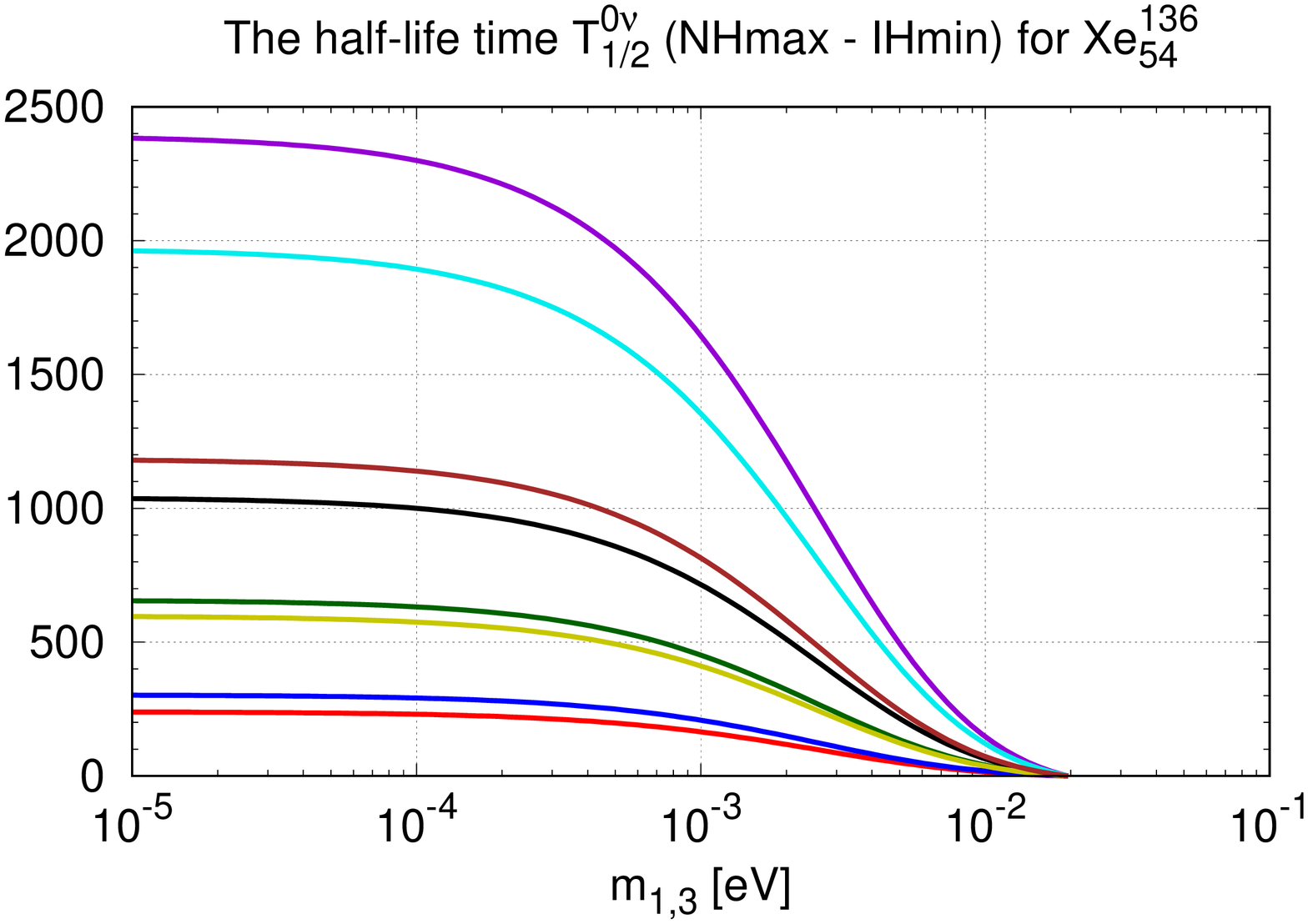}
\caption{half-lives and half-life differences for neutrinoless double beta decay in the standard approach and various nuclear matrix element calculations. The upper panel is the dependence of the maximal (solid) and minimal (dashed) half-life in the inverted ordering as a function of the solar neutrino mixing angle for $m_3=0$; the middle panel shows the half-lives related to the minimal $\langle m_{ee} \rangle$ in the inverted ordering (solid) and the maximal $\langle m_{ee} \rangle$ in the normal ordering (dashed) as a function of the solar neutrino mixing angle and smallest mass set to zero. The lower panel shows the half-life differences related to the minimal value in the inverted ordering and the maximal value in the normal ordering as a function of the smallest mass.\mbox{ }\vspace{.96543cm}}
\label{fig:diff}
\end{figure}

Of particular interest is the minimal value of the effective mass in the inverted ordering, $\langle m_{ee} \rangle^{\rm IH}_{\rm min} \simeq \sqrt{\Delta m^2_{\rm a}} \, \cos^2 \theta_{13} \, \cos 2 \theta_{12}$. It depends significantly on $\theta_{12}$ and JUNO will be particularly helpful in fixing this uncertainty \cite{Dueck:2011hu}. Another quantity of interest is the difference of this value with the maximum (zero $m_1$) value of the effective mass in the normal ordering, $\langle m_{ee} \rangle^{\rm NH}_{\rm max} \simeq \sqrt{\Delta m^2_{\rm a}} \, \sin^2 \theta_{13} + \sqrt{\Delta m^2_{\rm s}} \cos^2 \theta_{13} \, \sin^2 \theta_{12}$. We see from \gfig{fig:obs} that as 
long as the smallest mass is 
below about 0.01 eV one can in principle distinguish the normal from the inverted mass ordering with double beta decay. 
Using the nuclear matrix element compilation from a recent review \cite{Engel:2016xgb} we display in \gfig{fig:diff} various aspects related to the above discussion.

The upper panel shows the maximal and minimal half-life of neutrinoless double beta decay as a function of the solar neutrino  mixing angle in the inverted mass ordering with $m_3 =0$. The $3\sigma$ range of $\sin^2 \theta_{12} = 0.25 \ldots 0.354$ implies a spread of $\langle m_{ee} \rangle^{\rm IH}_{\rm min}$ of about 1.7, which leads to almost a 
factor of 3 for the half-life. For all considered isotopes the nuclear uncertainty exceeds the one from the oscillation parameters, hence the curves overlap.

The middle panel of \gfig{fig:diff} shows the half-lives for 
$\langle m_{ee} \rangle^{\rm IH}_{\rm min}$ (solid) and $\langle m_{ee} \rangle^{\rm NH}_{\rm max}$ (dashed) as a function of $\sin^2 \theta_{12}$ for vanishing smallest mass. In this case $\langle m_{ee} \rangle^{\rm NH}_{\rm max}$ is much smaller 
than $\langle m_{ee} \rangle^{\rm IH}_{\rm min}$, 
so that large half-life differences are present. Also, the dependence of $\langle m_{ee} \rangle^{\rm NH}_{\rm max}$ on $\theta_{12}$ is weaker than that of $\langle m_{ee} \rangle^{\rm IH}_{\rm min}$. 
The nuclear uncertainty for $^{130}$Te and $^{136}$Xe is larger than for $^{76}$Ge (3.11 and 3.16 versus 2.18 for the ratio of largest to smallest matrix element). 
For the two former isotopes it could thus happen in the plot that the half-life for 
$\langle m_{ee} \rangle^{\rm IH}_{\rm min}$ looks actually larger than the one for 
$\langle m_{ee} \rangle^{\rm NH}_{\rm max}$. 

Finally, the lower panel of \gfig{fig:diff} shows the half-life difference for 
$\langle m_{ee} \rangle^{\rm IH}_{\rm min}$ and $\langle m_{ee} \rangle^{\rm NH}_{\rm max}$ as a function of the smallest mass when all oscillation parameters are fixed to their best-fit values. We see that for smallest masses below a few times $10^{-4}$ eV there is no change with respect to the zero mass case. For smallest masses about $10^{-3}$ eV the half-life differences  decrease by about 40\%. \\

\begin{figure}[t]
\begin{center}
\includegraphics[width=6cm, height=7.7cm,angle=-90]{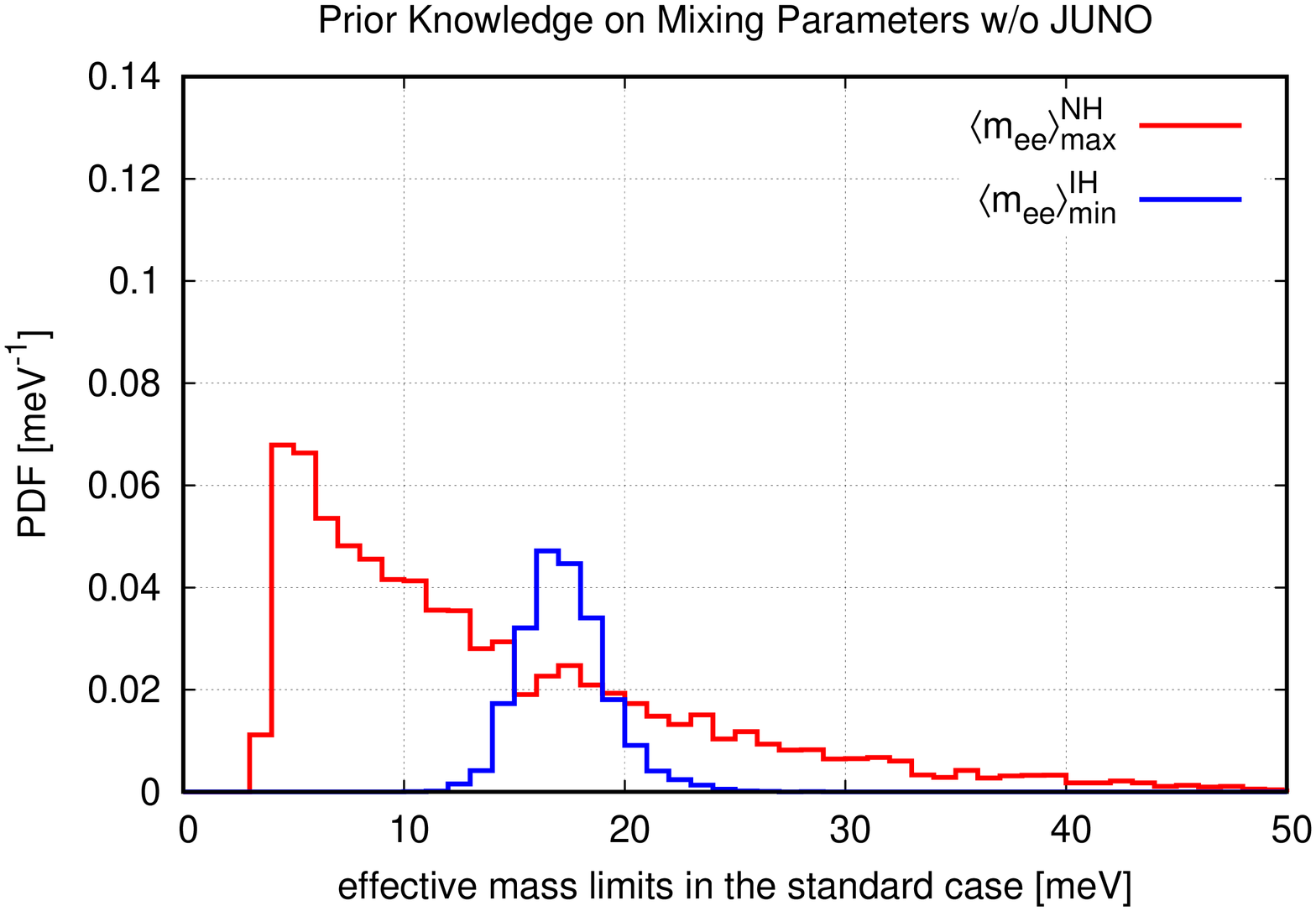} 
\includegraphics[width=6cm, height=7.7cm,angle=-90]{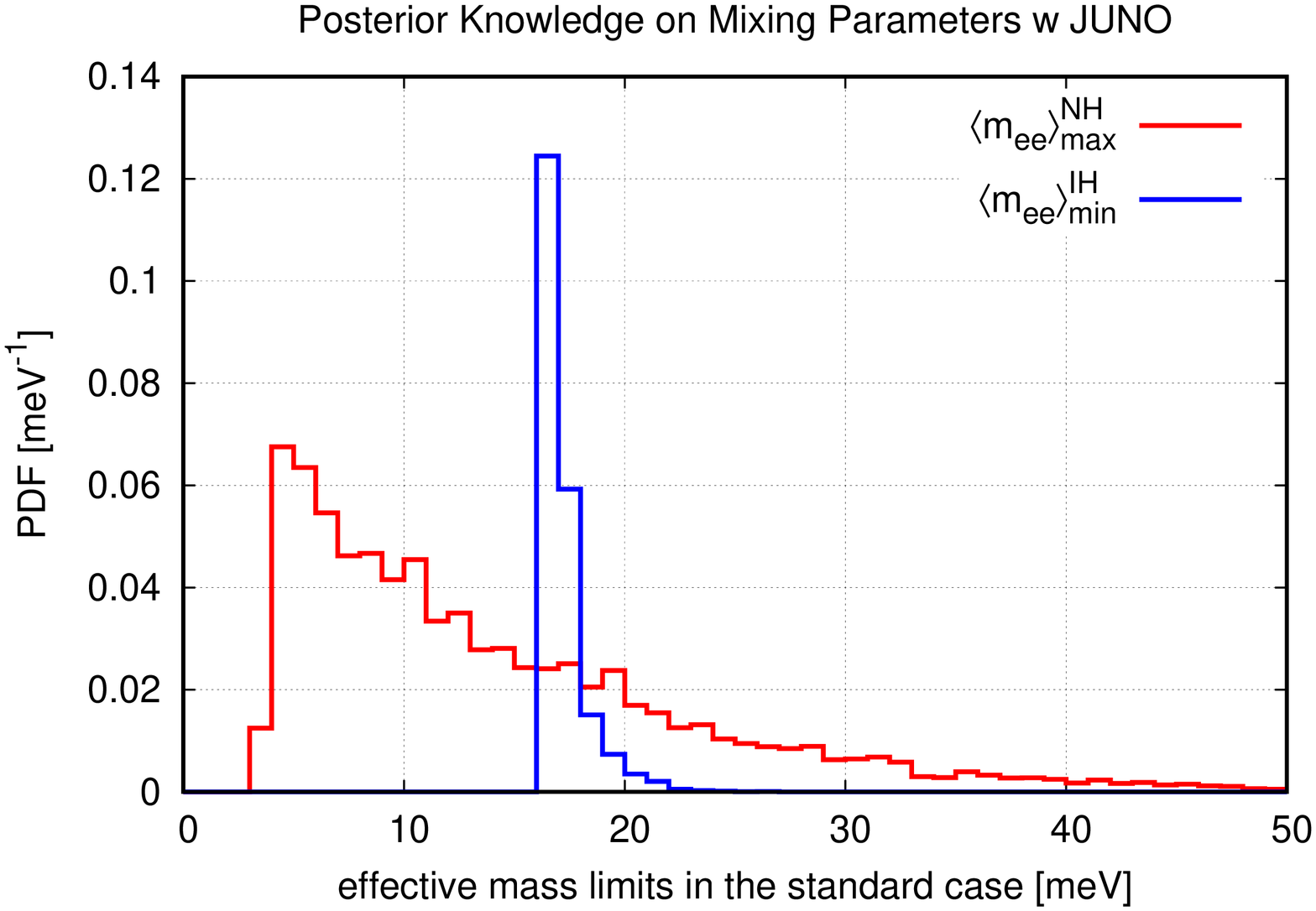} 
\caption{The distribution of the effective mass (maximal value for the normal and minimal value for the inverted ordering)
in the standard scenario using constraints on 
the sum of neutrino masses from cosmology and the current (left) or 
prospective future (right) precision of the oscillation parameters.}
\label{fig:NH-IH-standard}
\end{center}
\end{figure}\vspace{-.5cm}
\begin{figure}[h!]
\begin{center}
\includegraphics[width=6cm, height=7.7cm,angle=-90]{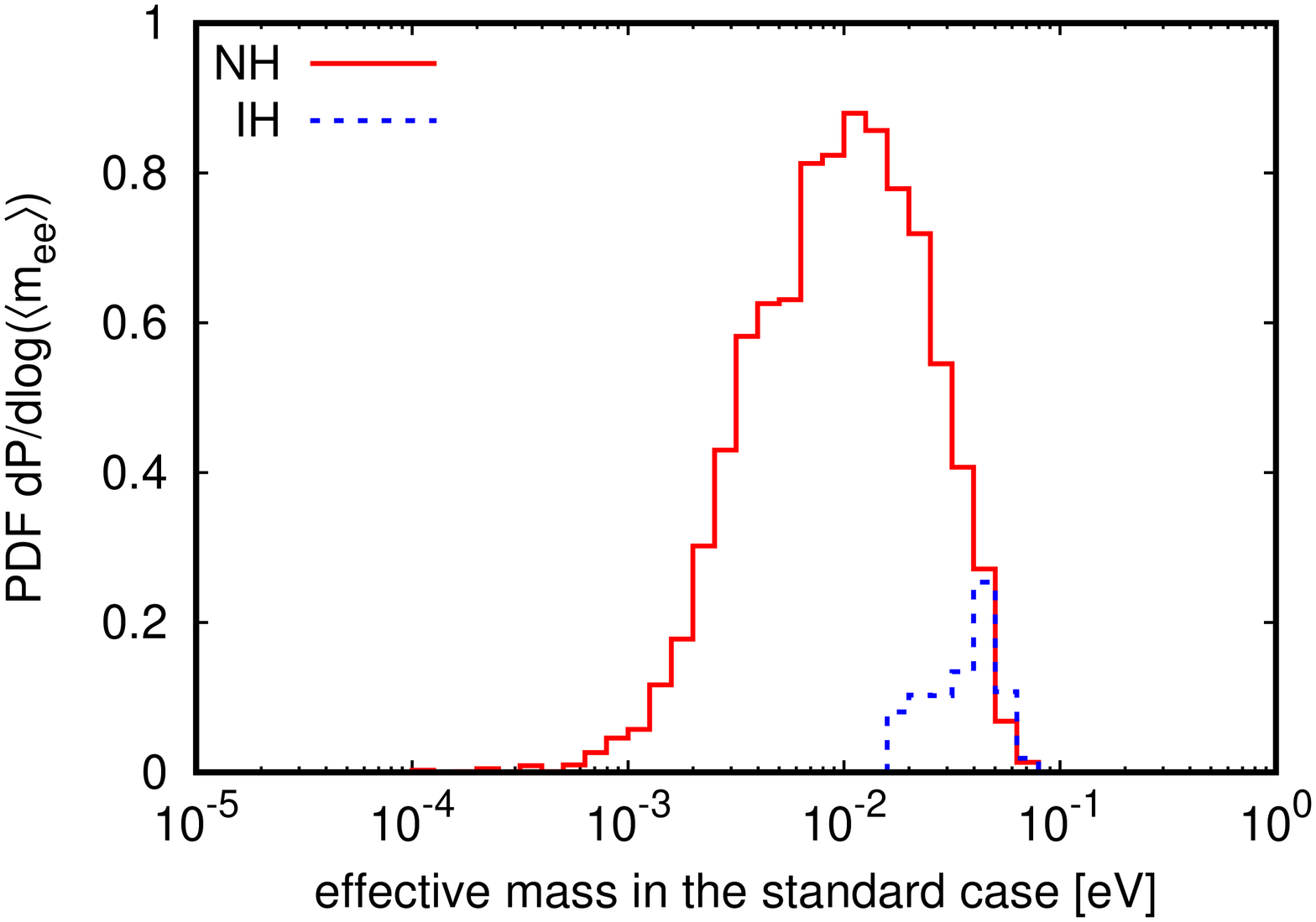} 
\caption{The distribution of the effective mass in the standard scenario using constraints on 
the sum of neutrino masses from cosmology and the current precision of the oscillation parameters.\mbox{}\\}
\label{fig:meff}
\end{center}
\end{figure}

Following the procedure outlined in Section \ref{sec:proc}, we show in \gfig{fig:NH-IH-standard}  
the probability distribution of the maximal value of the 
effective mass in the normal ordering, $\langle m_{ee}\rangle^{\rm NH}_{\rm max}$, 
and of the minimal value in the inverted ordering, 
$\langle m_{ee}\rangle^{\rm IH}_{\rm min}$. 
\begin{figure}[h!]
\centering
\includegraphics[width=6cm, height=7.7cm, angle=-90]{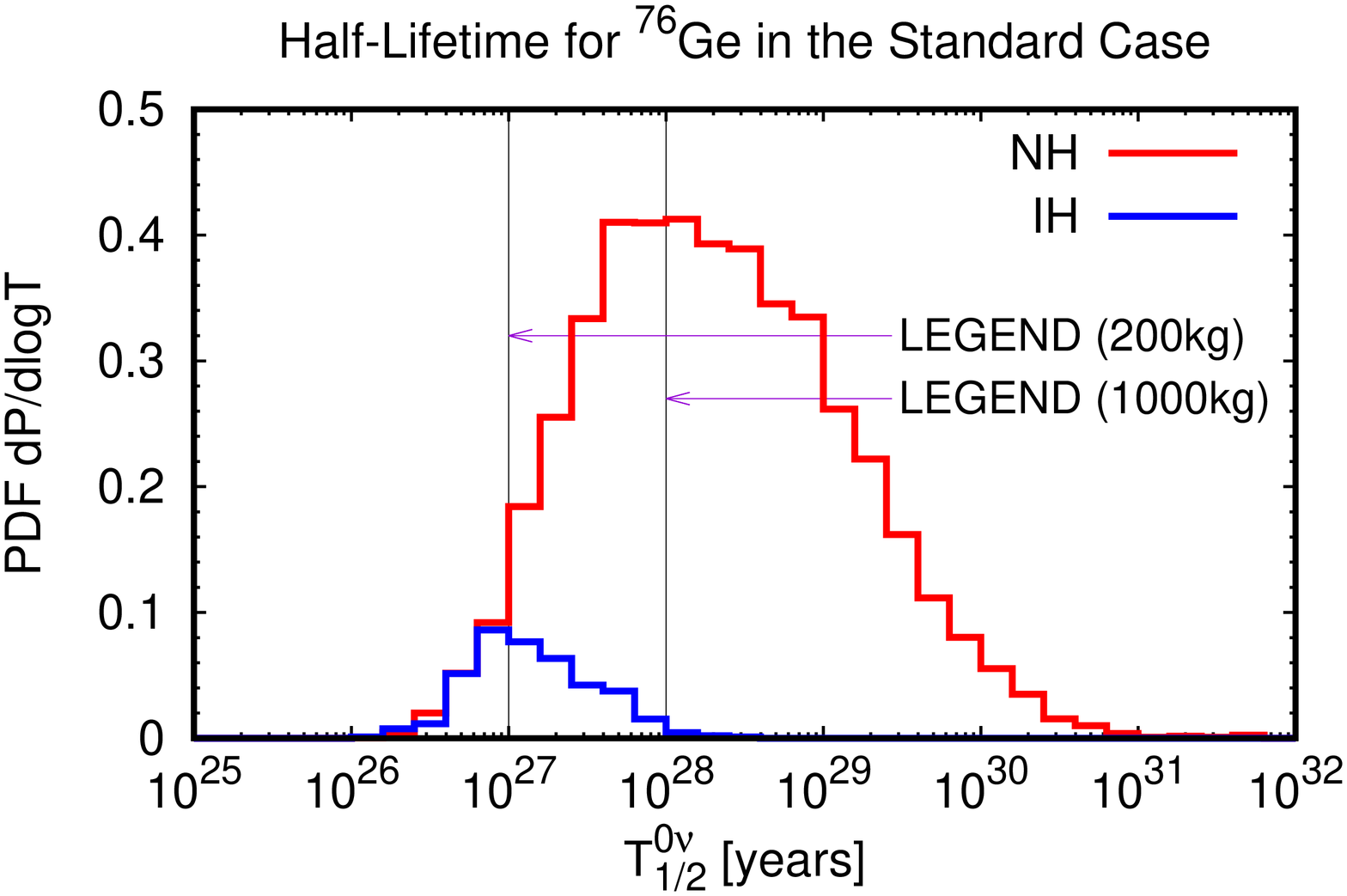}
\includegraphics[width=6cm, height=7.7cm, angle=-90]{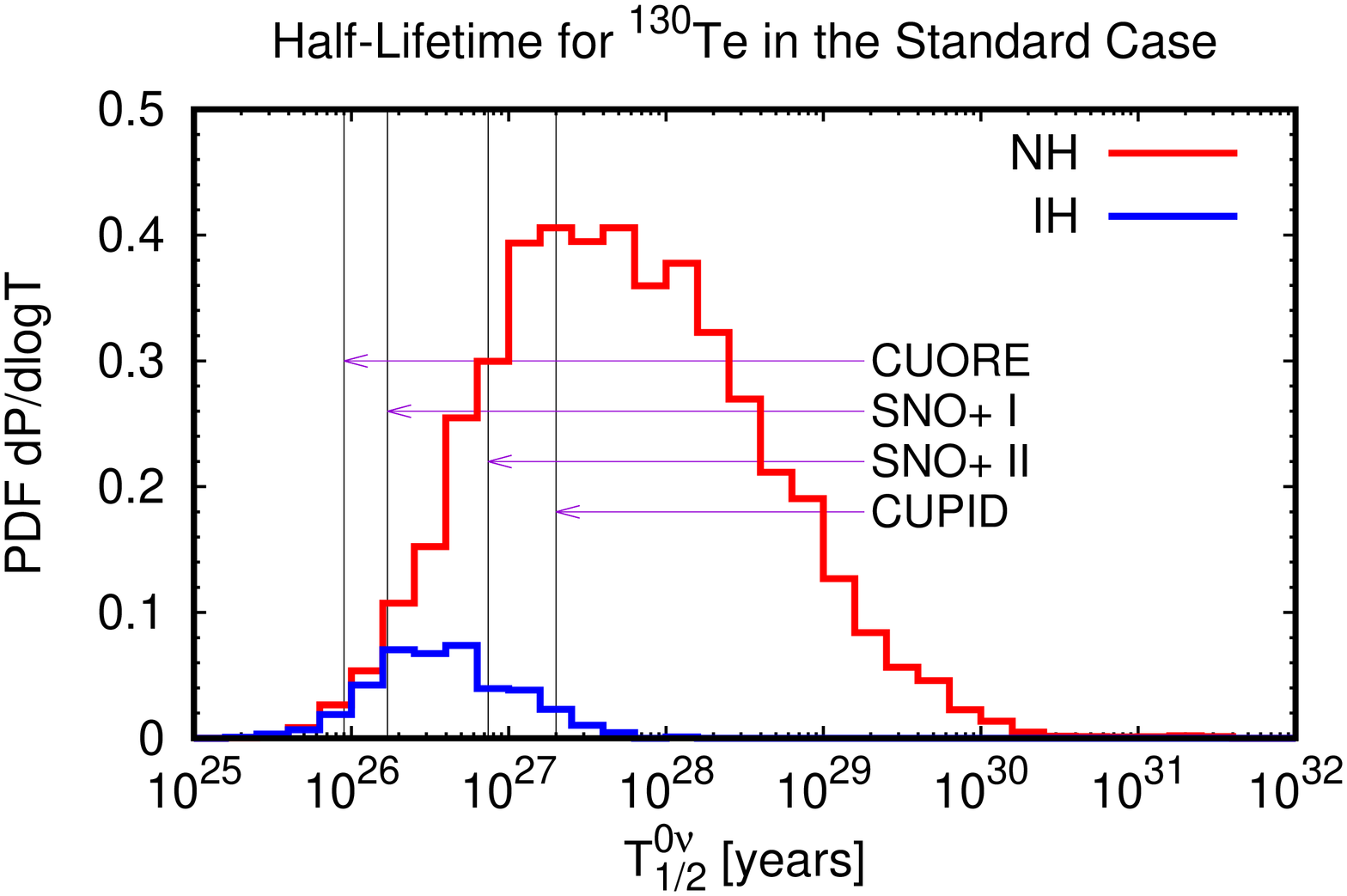}
\includegraphics[width=6cm, height=7.7cm, angle=-90]{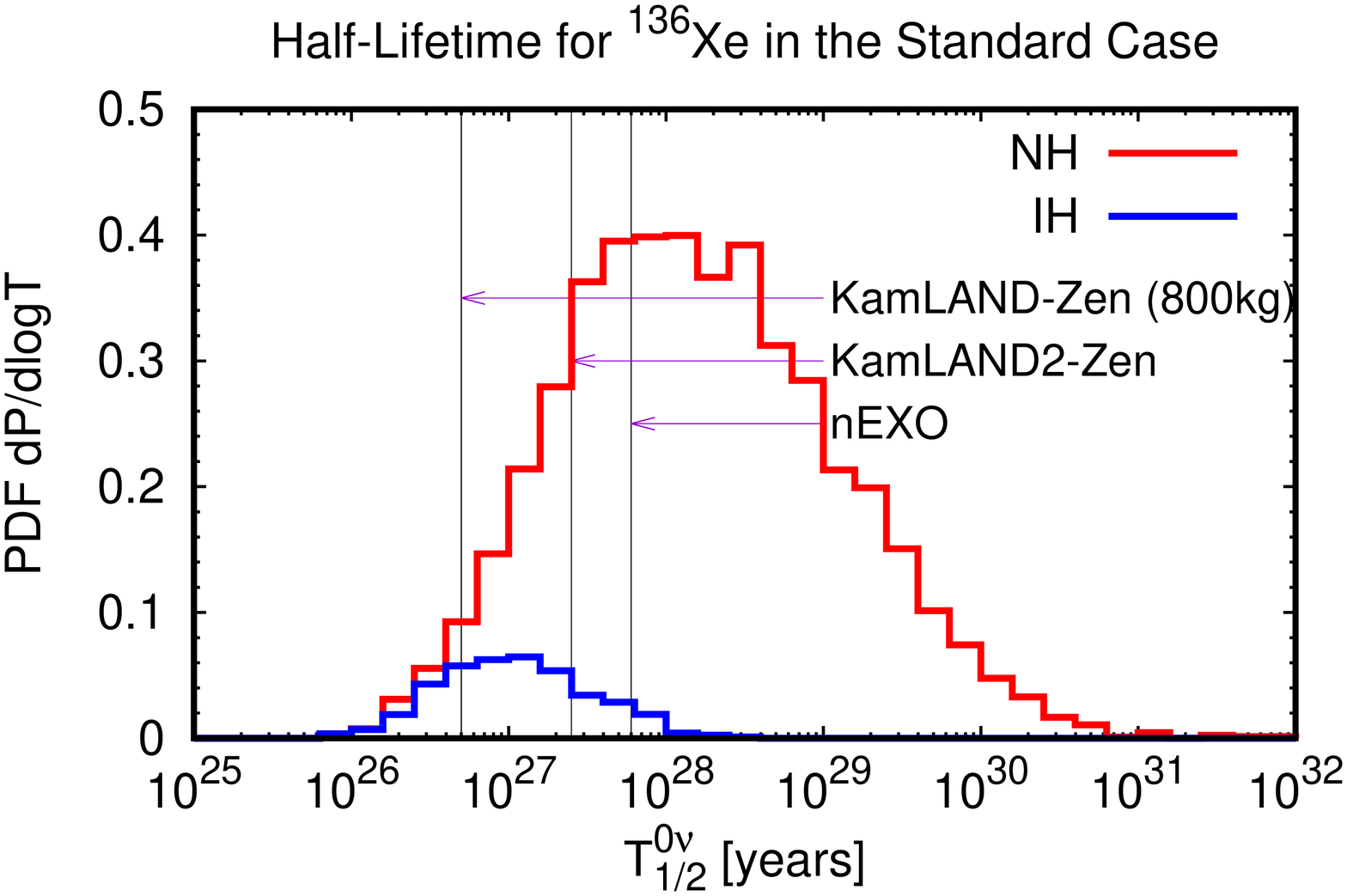}
\caption{The predicted distribution of neutrinoless double beta decay half-lives in the standard approach.}
\label{fig:sampleThalf}
\end{figure}
The left plot shows the situation for the current ranges of the oscillation parameters, the right plot 
for the situation after JUNO has determined in particular the solar neutrino mixing angle $\theta_{12}$ with remarkable precision. 
As discussed in Section \ref{sec:nu}, the  
uncertainty on $\langle m_{ee} \rangle^{\rm IH}_{\rm min}$, which is proportional to $\cos 2 \theta_{12}$, 
is significantly reduced  in this case \cite{Dueck:2011hu,Ge:2015bfa}. Indeed, the distribution of $\meeIHmin$ shown in \gfig{fig:NH-IH-standard} becomes significantly narrower
after including the JUNO constraint. However, the distribution of $\meeNHmax$
is not that sensitive to the solar angle. 
The result for the distribution of the effective mass 
(see also \cite{Bergstrom:2012nx,Benato:2015via,DellOro:2015kys,Zhang:2015kaa,Ge:2016tfx}) 
is shown in \gfig{fig:meff}. We see that the inverted ordering is centered around larger values than the normal ordering, but the normal ordering has significant probability for an effective mass above 0.01 eV. The relative smallness of the curves for the inverted ordering originates from the combined preference of $12:1$ for the normal mass ordering. \\

Using the result for the effective mass, \gfig{fig:sampleThalf} shows the predicted distributions of neutrinoless double
beta decay half-life which are obtained from convoluting the probability
distribution of the effective mass in \gfig{fig:meff} and the correlated 
Gaussian distribution of NMEs \cite{Faessler:2008xj,Faessler:2013hz} as summarized 
in \gtab{tab:NME}. 
It allows to compare our results with the ones from Refs.\ \cite{Agostini:2017jim,Caldwell:2017mqu}. For instance, a half-life limit of $1\times 10^{27}$ years as achievable by LEGEND-200 with $^{76}$Ge would cover 3.77\% of the expected normal ordering range and 39.2\% of the inverted ordering one\footnote{The experimental half-life ranges we will use for illustration correspond to the $3\sigma$ sensitivity after 4 years of running.}. CUPID, with a sensitivity of $2\times10^{27}$ years using $^{130}$Te, could cover 32.7\% of the normal and 93.5\% of the inverted ordering.  
The nEXO experiment, with a possible limit of $6\times 10^{27}$ years using $^{136}$Xe, could cover 33.6\% of the expected normal ordering range and 92.5\% of the inverted ordering one. The coverage of the different experimental projects for the two mass orderings is summarized in \gtab{tab:res}. For completeness, we also give in  \gtab{tab:res1} the necessary half-lives to exclude 95\% of the half-life values on the normal and inverted ordering. 


\section{\label{sec:st}Neutrinoless Double Beta Decay with Light Sterile Neutrinos}

There are long-standing and hard-to-kill hints towards the presence of light sterile neutrinos \cite{Gariazzo:2015rra}. What is relevant for neutrinoless double beta decay is the mass-squared difference $\Delta m^2_{41}$ and the mixing of the fourth sterile state $\nu_4$ 
with electron neutrinos, $U_{e4}$. We use the fit results from 
Ref.\ \cite{Gariazzo:2017fdh}, namely the ``{PrGlo17}'' curve in the Fig.\ 9 therein,
and assume that the smallest neutrino mass is zero (hence we can use to a reasonable precision the cosmology fit results we used for the standard case above), thus we have $m_4 = \sqrt{\Delta m^2_{41}}$. These are valid assumptions as long as the smallest mass is below a few times $10^{-2}$ eV. 
We do not take into account the preference towards the normal mass ordering from the cosmology fit \cite{Hannestad:2016fog}. To be specific, the best-fit values
are $\Delta m^2_{41} \approx 1.75\,\mbox{eV}^2$ and $|U_{e4}|^2 \approx 0.02$.

\begin{figure}[t!]
\centering
\includegraphics[height=10cm,angle=-90]{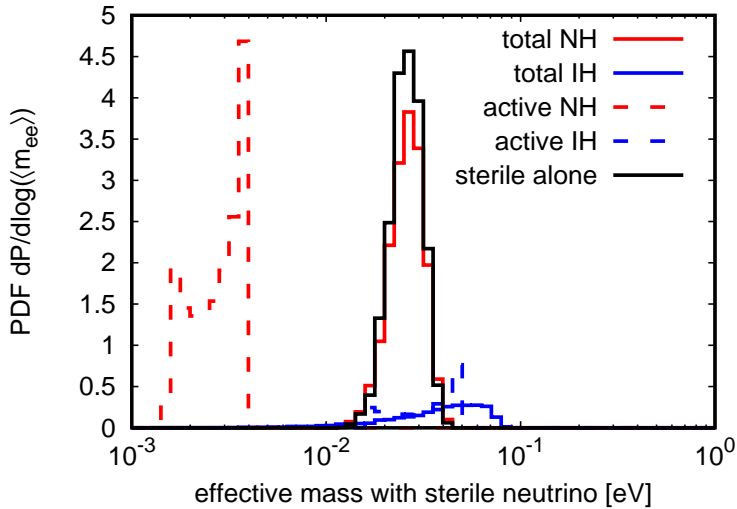}
\caption{The predicted distribution of the effective
mass $\langle m_{ee} \rangle$ in the presence of a light sterile neutrino and zero smallest mass. }
\label{fig:mee-Sterile}
\end{figure}

The presence of a sterile neutrino contributes an extra term to the effective mass: 
\begin{equation} \label{eq:st}
  \mee^{\rm sterile}
=
\left|U_{e1}^2 \,m_1 + U_{e2}^2 \,m_2 \,e^{i \alpha} + U_{e3}^2 \,m_3 \,e^{i \beta} +   m_4 \,U^2_{e4} \, e^{i \gamma}
\right| \,.
\end{equation}
The extra mixing matrix element $U_{e4}$ contains a new Majorana CP phase $\gamma$ which could
have a significant effect on the predicted effective mass. 
As has been noted in Ref.\ \cite{Barry:2011wb} (see also \cite{Li:2011ss,Girardi:2013zra,Giunti:2015kza}), the typical fit results for $\Delta m^2_{41}$ and $U_{e4}$ imply that the sterile contribution $|m_4 \,U^2_{e4}|$ is of the same order as the active contribution (the first three terms in (\ref{eq:st})) if the active neutrinos are inversely ordered. Hence the total effective mass can vanish in this case. 
If the active neutrinos are normally ordered and the neutrino mass is small, the sterile contribution is much larger and the total effective mass can not vanish. Therefore, the phenomenology of double beta decay and the mass ordering is fundamentally opposite to the standard case, and consequently the physics potential of future experiments changes drastically.

\gfig{fig:mee-Sterile} shows the probability distribution of $\mee$ in the presence
of a light eV-scale sterile neutrino. 
Indeed, the contribution from sterile neutrino alone is comparable with the active contribution in the inverted ordering case. The 
active and sterile neutrino contributions add randomly for the inverted ordering, broadening the predicted distribution. 


\begin{figure}[t]
\centering
\includegraphics[width=6cm, height=7.7cm, angle=-90]{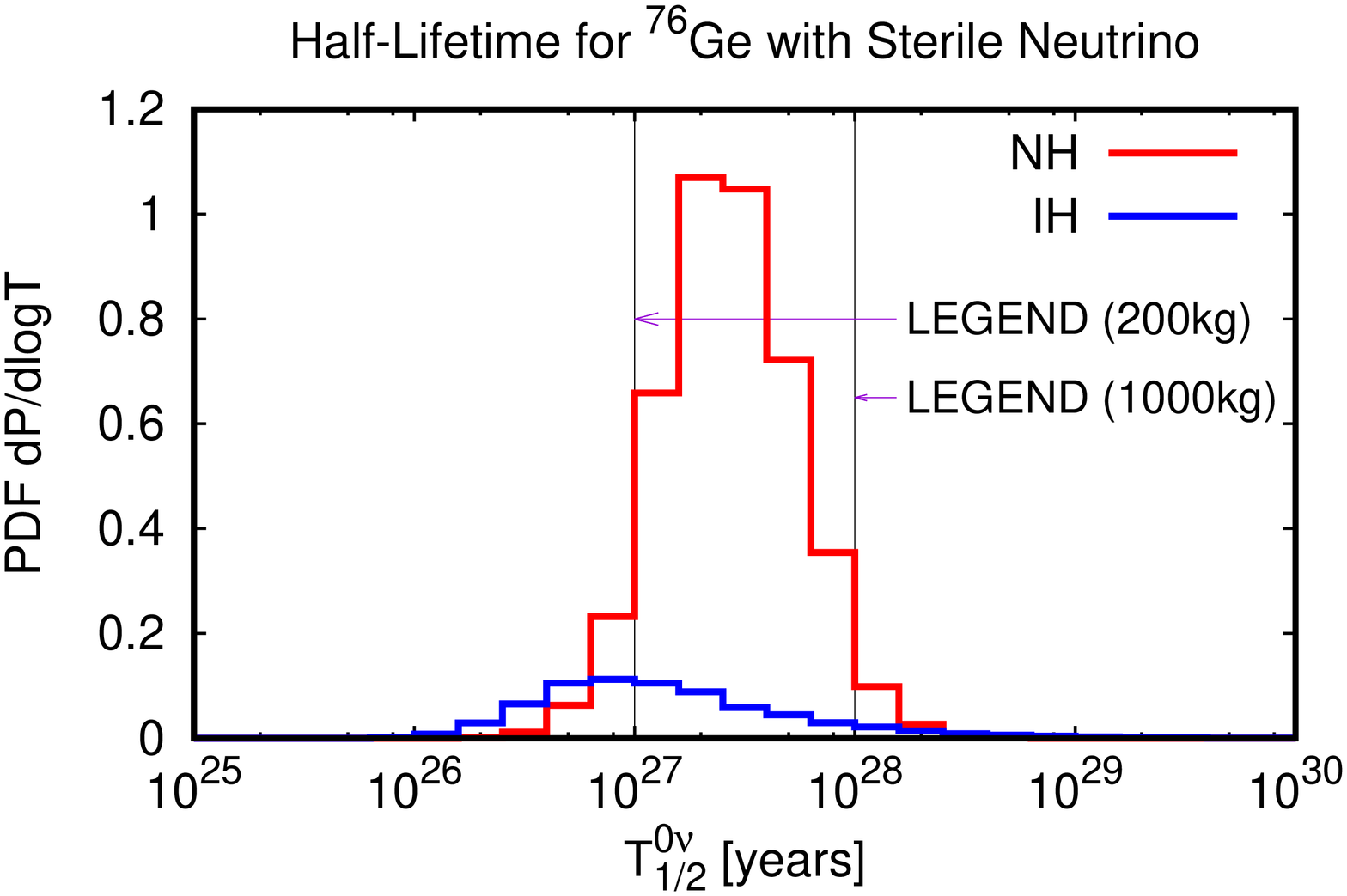}
\includegraphics[width=6cm, height=7.7cm, angle=-90]{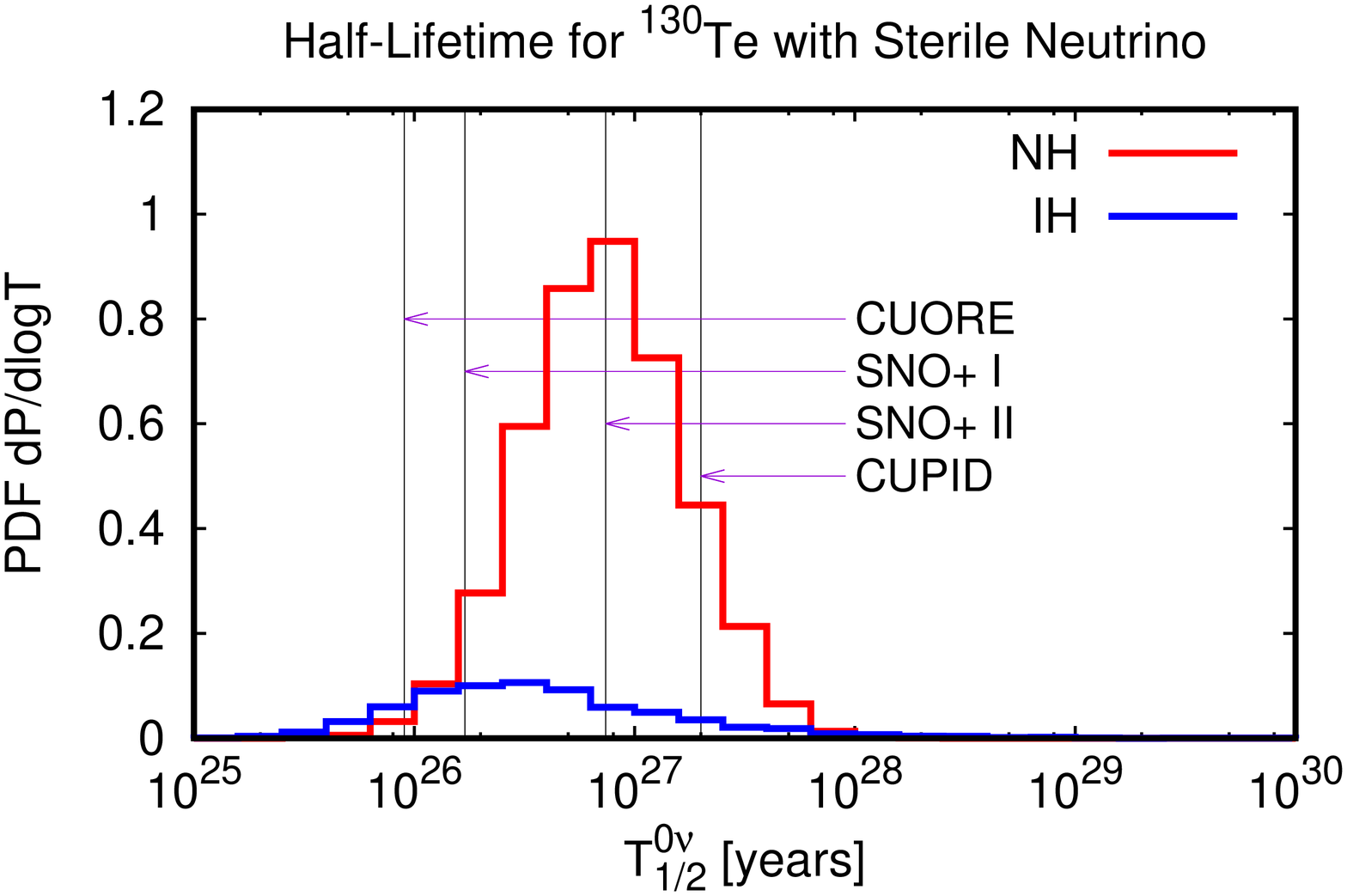}
\includegraphics[width=6cm, height=7.7cm, angle=-90]{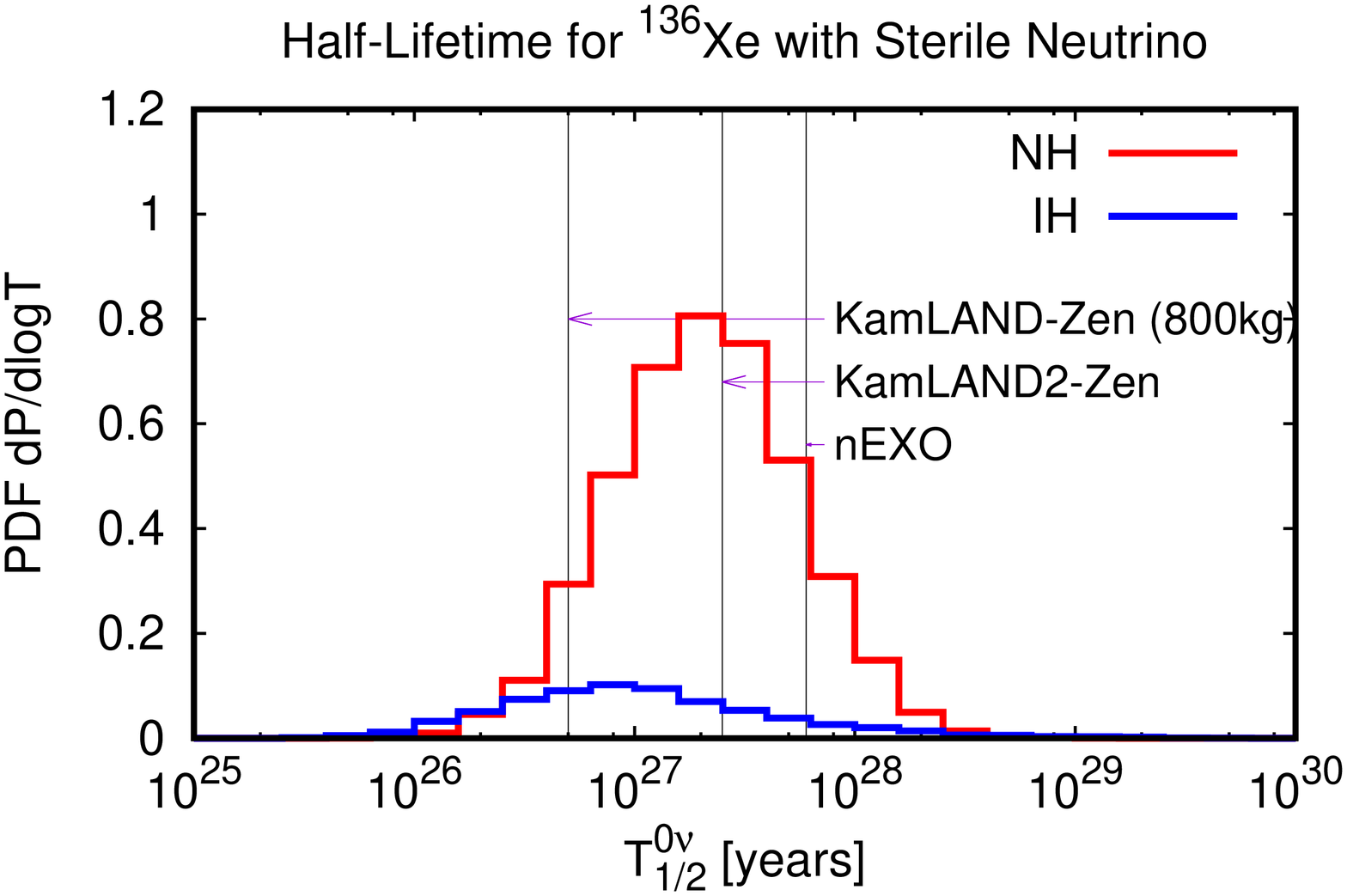}
\caption{The predicted distribution of neutrinoless double beta decay half-life 
in the presence of a sterile neutrino.}
\label{fig:sampleThalf-Sterile}
\end{figure}

\gfig{fig:sampleThalf-Sterile} shows the predicted half-life distribution, to be compared to the standard 3-neutrino approach from \gfig{fig:sampleThalf}. 
Due to the contribution from the sterile neutrino, the predicted half-lives for the inverted ordering are shifted towards larger half-lives. Since the effective mass 
spans also a wider range for this case, the distribution is also wider.
For instance, a half-life limit of $1\times 10^{27}$ years as achievable by LEGEND-200 would cover 7.63\% of the expected normal ordering range and 45.4\% of the inverted ordering one. CUPID, with a sensitivity of $2\times10^{27}$ years, could cover 88.4\% of the normal and 88.2\% of the inverted ordering. The nEXO experiment, with a possible limit of $6\times 10^{27}$ years, could over 86.4\% of the expected normal ordering range and 87.7\% of the inverted ordering one. 
The coverage of the different experimental projects for the two mass orderings is summarized in \gtab{tab:res}. The necessary half-lives to cover 95\% of the expected 
half-lives are summarized in \gtab{tab:res1}.

\section{\label{sec:LR}Neutrinoless Double Beta Decay in Left-Right Symmetric Theories}

Double beta decay can be generated by many possible mechanims  beyond the standard Majorana neutrino exchange diagram \cite{Rodejohann:2011mu}. Typically, the parameters relevant for the non-standard diagrams are not directly related to neutrino mass. A popular and attractive exception exists within left-right symmetric theories \cite{Pati:1974yy,Mohapatra:1974gc,Senjanovic:1975rk,Senjanovic:1978ev} in case type II dominance holds \cite{Tello:2010am}. 
Canonical left-right symmetric theories automatically contain a type I 
and a type II seesaw mechanism for neutrino mass. In case the type II seesaw term for neutrino mass \cite{Magg:1980ut,Mohapatra:1980yp,Lazarides:1980nt,Schechter:1980gr} dominates, and the usual discrete left-right symmetry applies, then the right-handed neutrino mass matrix is diagonalized by the PMNS matrix, and the heavy neutrino masses are directly proportional to the light ones. See for instance Refs.\ \cite{Tello:2010am,Barry:2013xxa,Ge:2015yqa} for more details. 

\begin{figure}[t!]
\centering
\includegraphics[width=6cm, height=7.7cm, angle=-90]{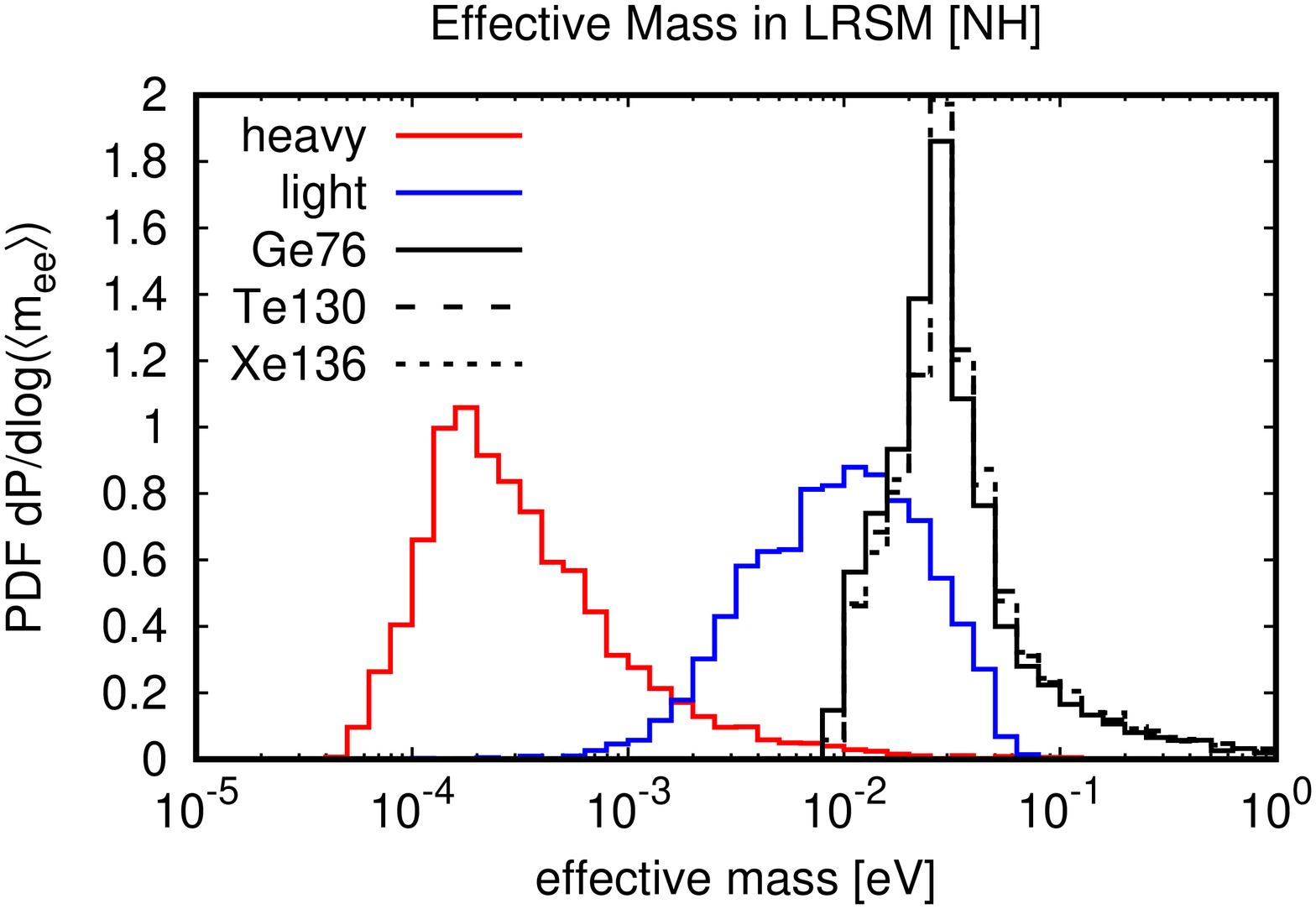}
\includegraphics[width=6cm, height=7.7cm, angle=-90]{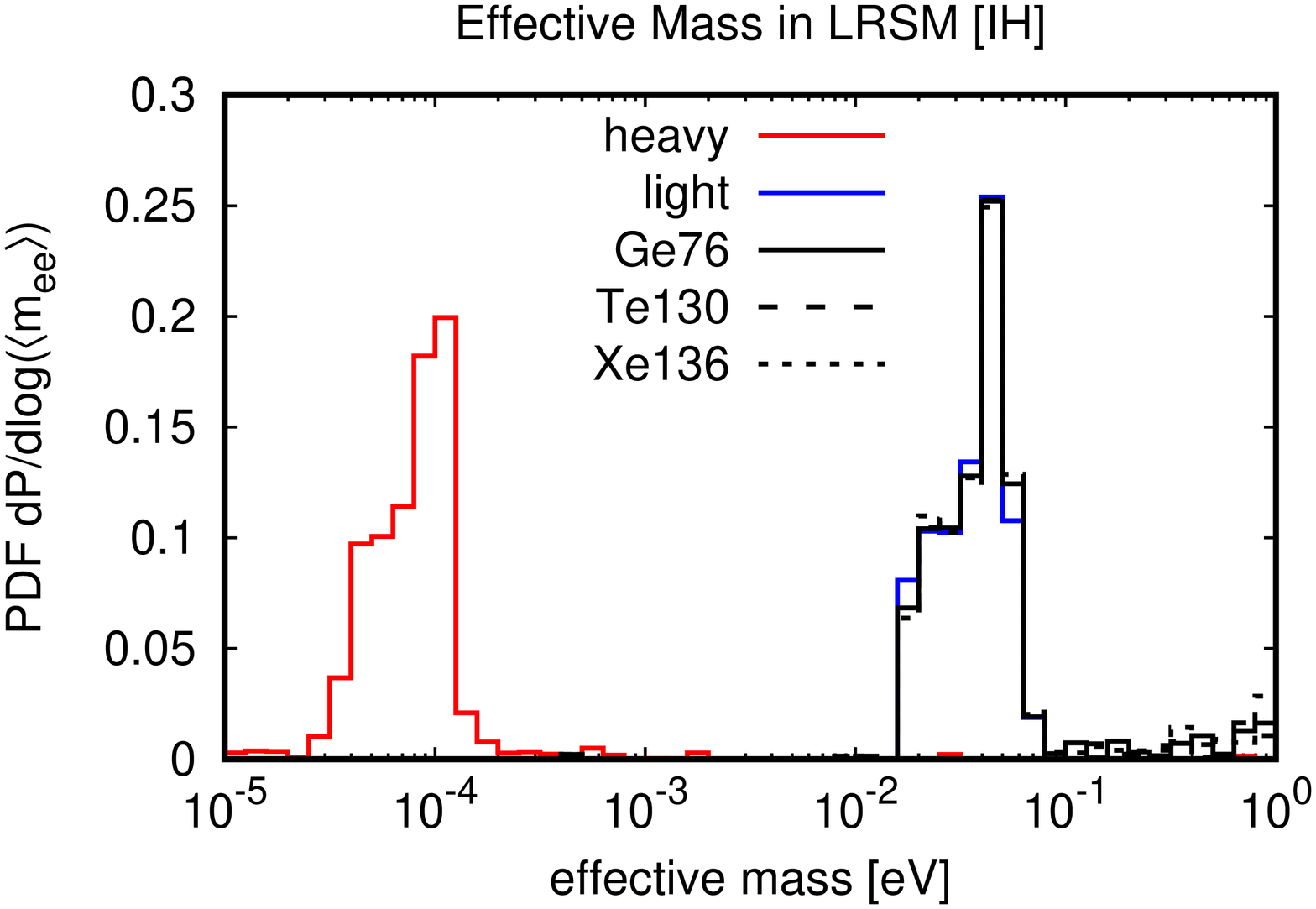}
\caption{The predicted distribution of the neutrinoless double beta decay effective masses 
for left-right symmetric theories with type II dominance ($g_R/g_L = \frac 23$, $M_{W_R} = 2.5$ TeV and a 
heaviest heavy neutrino mass of 1 TeV). 
The red (blue) lines show $\mee^{\rm heavy}$ $(\mee^{\rm light})$ from Eq.\ (\ref{eq:lr}), the black lines show 
the total $\mee^{\rm LR}$, which is isotope-dependent due to the nuclear matrix elements ratio in Eq.\ (\ref{eq:lr}). }
\label{fig:mee-LRSM}
\end{figure}

The new diagram under consideration is then the exchange of TeV-scale heavy neutrinos with TeV-scale right-handed $W_R$ bosons. Due to the right-handed electrons in the final state the new diagram adds incoherently to the standard one, and 
the new "effective mass" is 
\begin{equation} \label{eq:lr}
\begin{array}{c}
   \mee^{\rm LR}\displaystyle
= \mee^{\rm light} + \frac {\mathcal M^{\rm H}}
        {\mathcal M^{\rm L}} \mee^{\rm heavy} \\ \displaystyle = 
  \mee^{\rm light}  + \frac {\mathcal M^{\rm H}}{\mathcal M^{\rm L}} 
m_e m_p 
\left|
  \left( \frac {g_R M_{W_L}}{g_L M_{W_R}} \right)^4
	\left( \left. \mee^{\rm light} \right|_{m_i \rightarrow \frac {1}{M_i}} \right)
\right| .
\end{array}
\end{equation}
%
%
\begin{figure}[t!]
\centering
\includegraphics[width=4cm,height=0.32\textwidth,angle=-90]{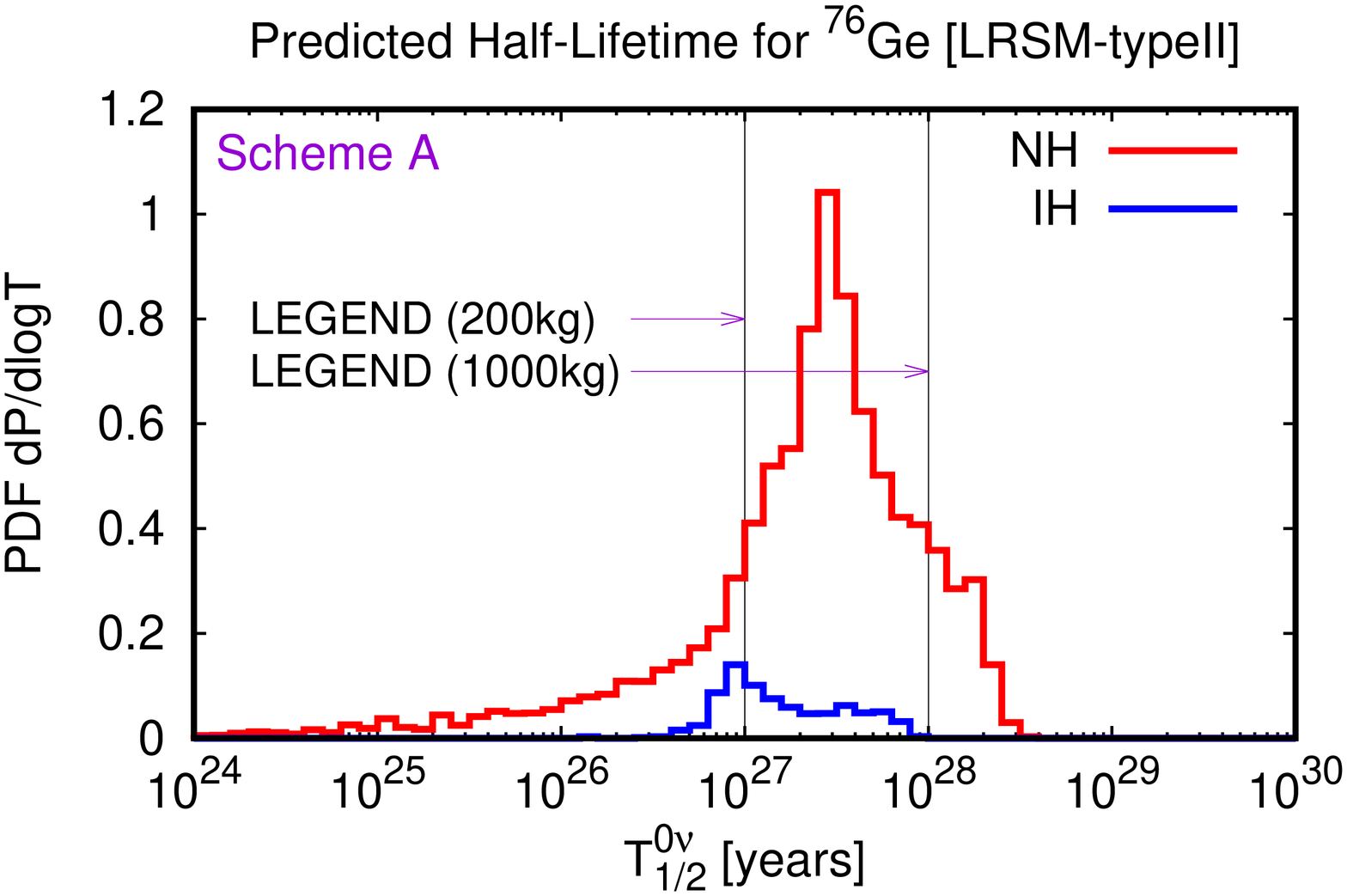}
\includegraphics[width=4cm,height=0.32\textwidth,angle=-90]{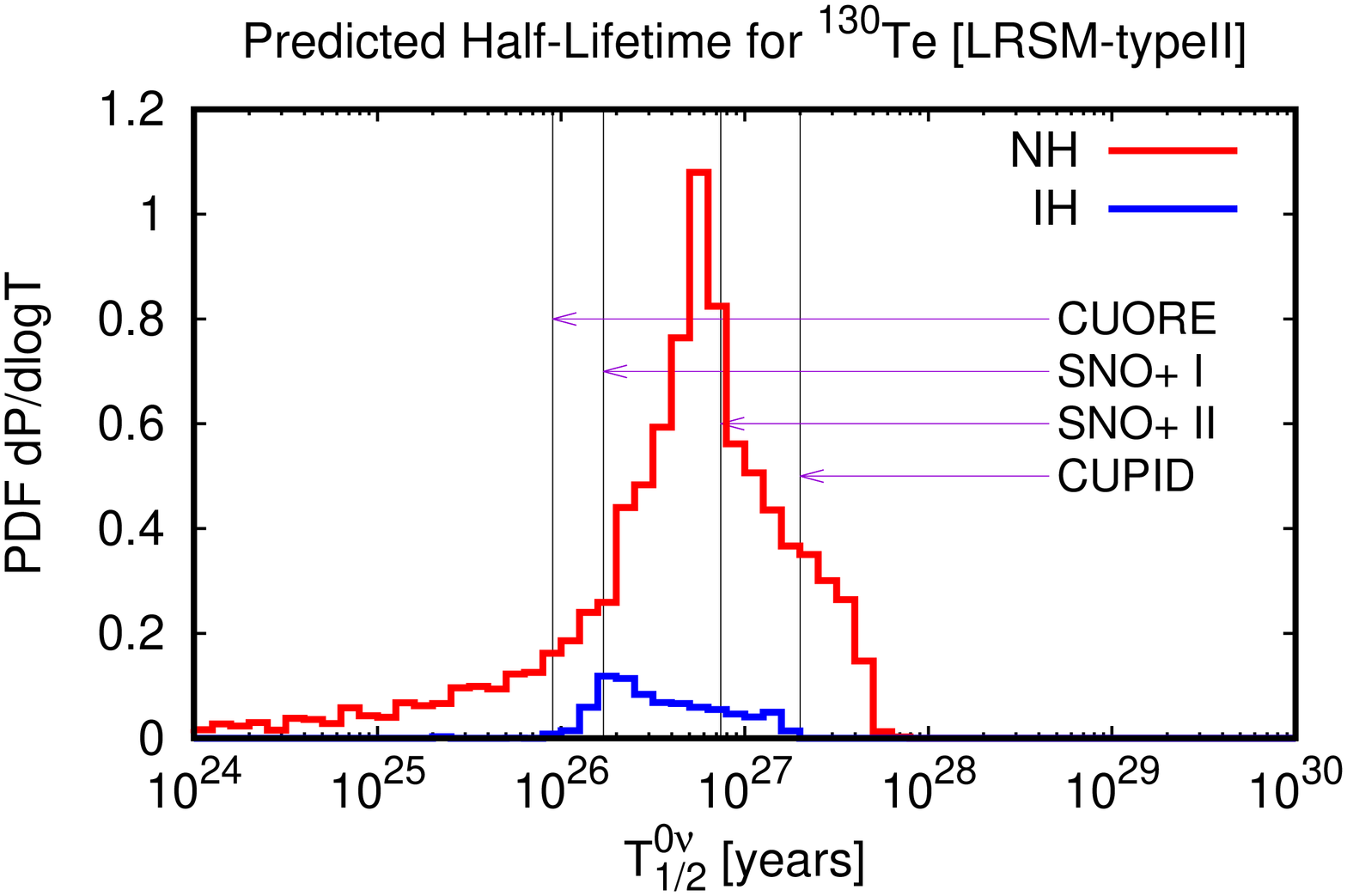}
\includegraphics[width=4cm,height=0.32\textwidth,angle=-90]{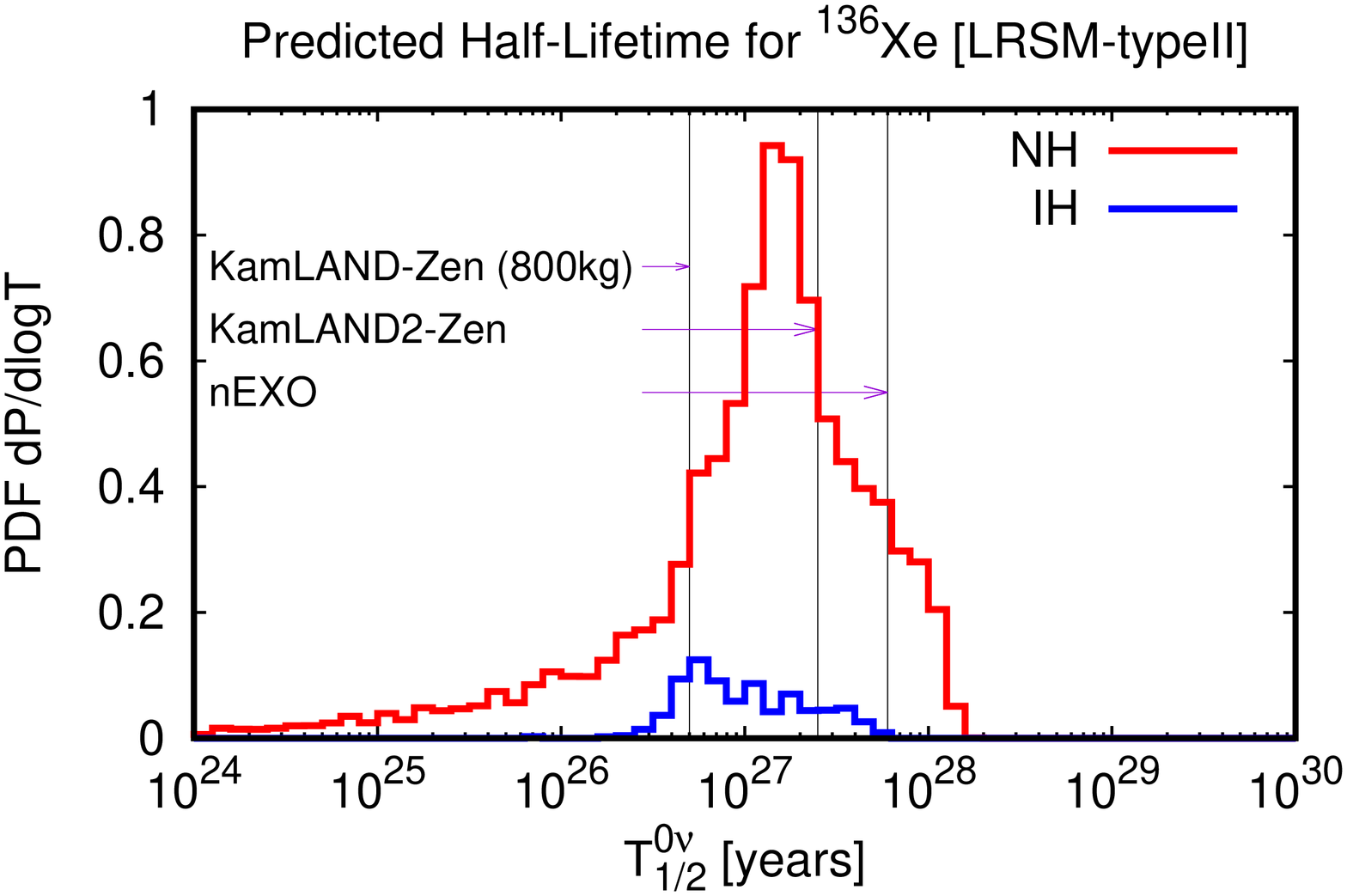}
\includegraphics[width=4cm,height=0.32\textwidth,angle=-90]{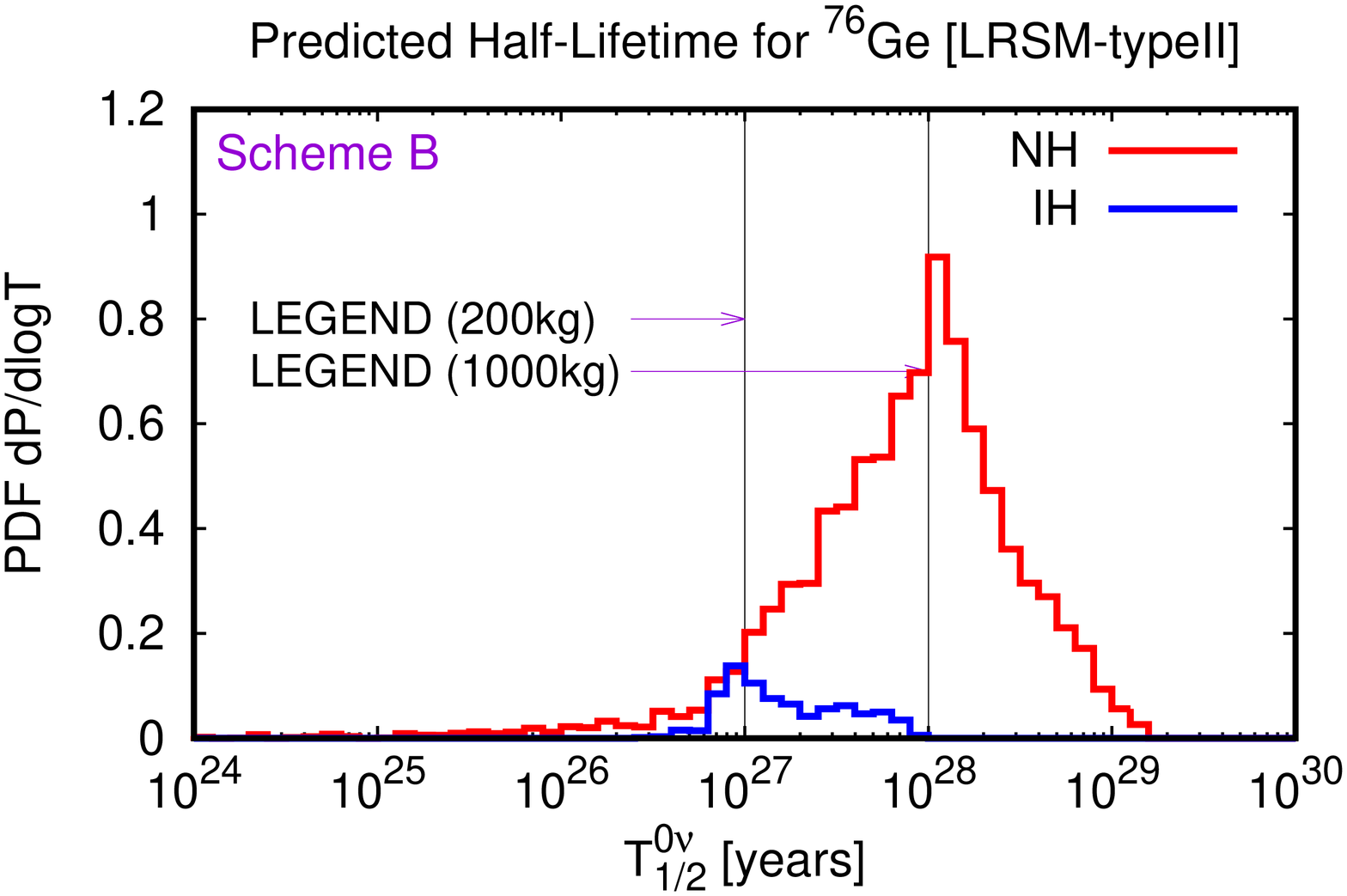}
\includegraphics[width=4cm,height=0.32\textwidth,angle=-90]{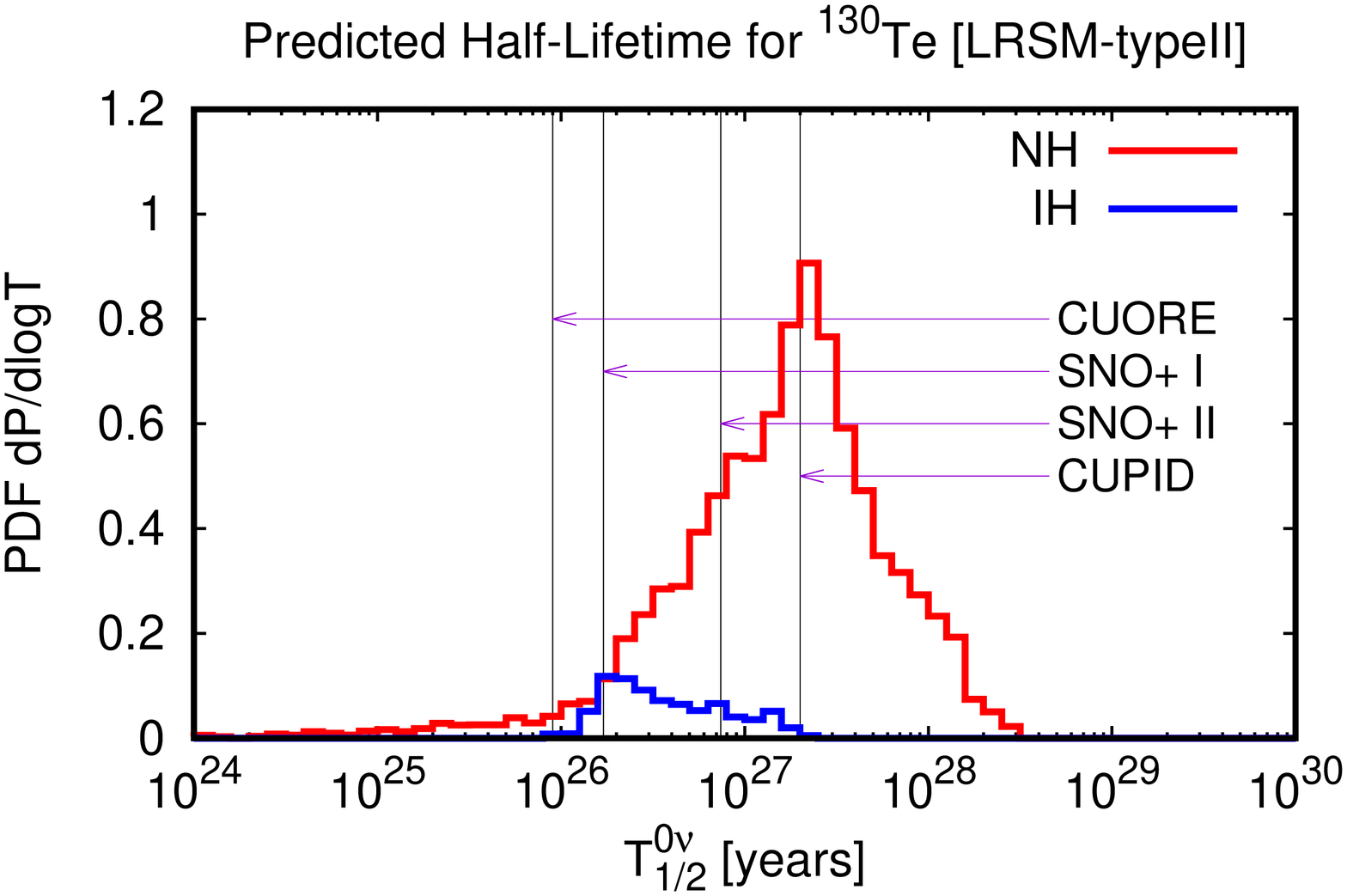}
\includegraphics[width=4cm,height=0.32\textwidth,angle=-90]{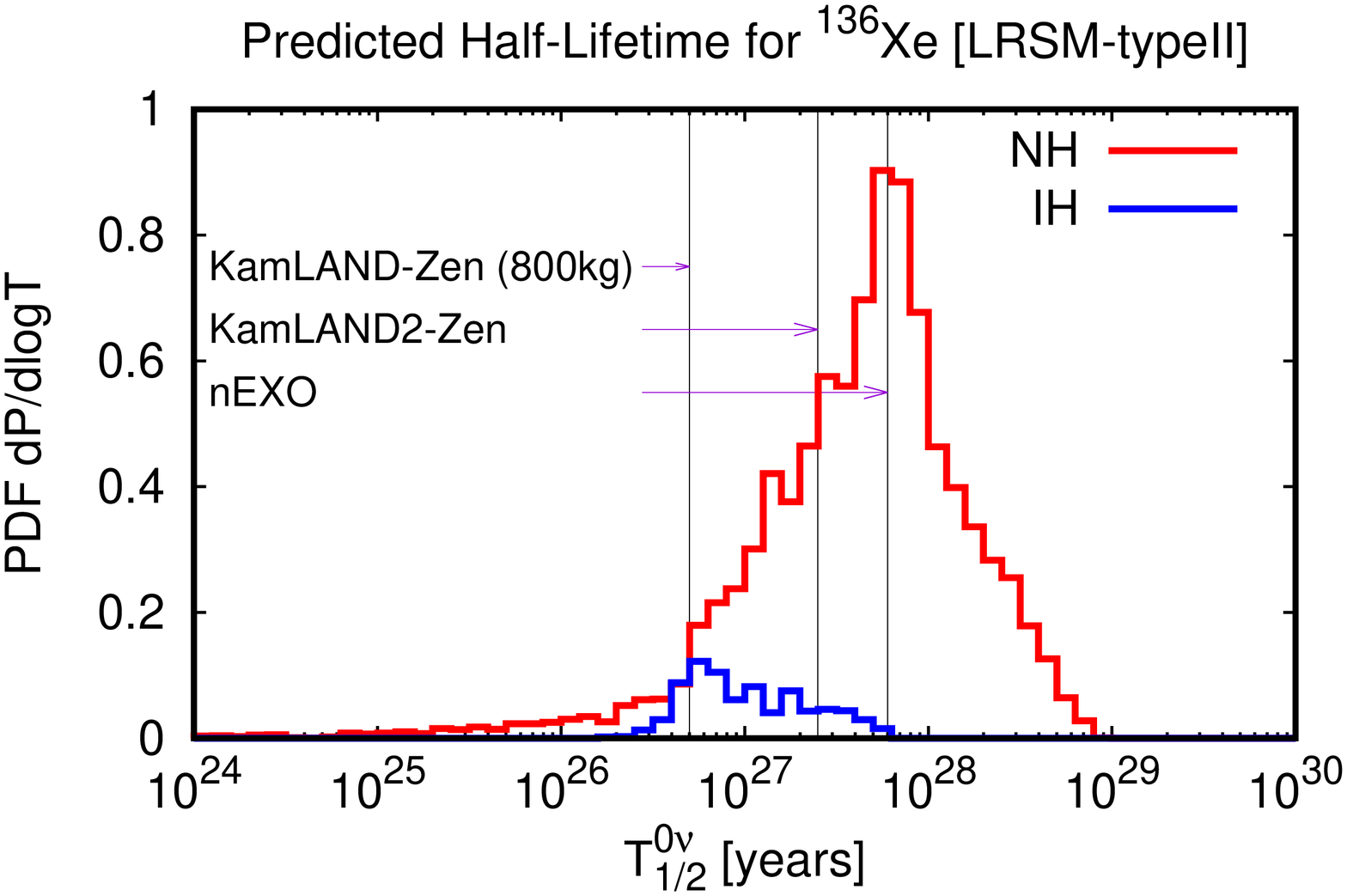}
\includegraphics[width=4cm,height=0.32\textwidth,angle=-90]{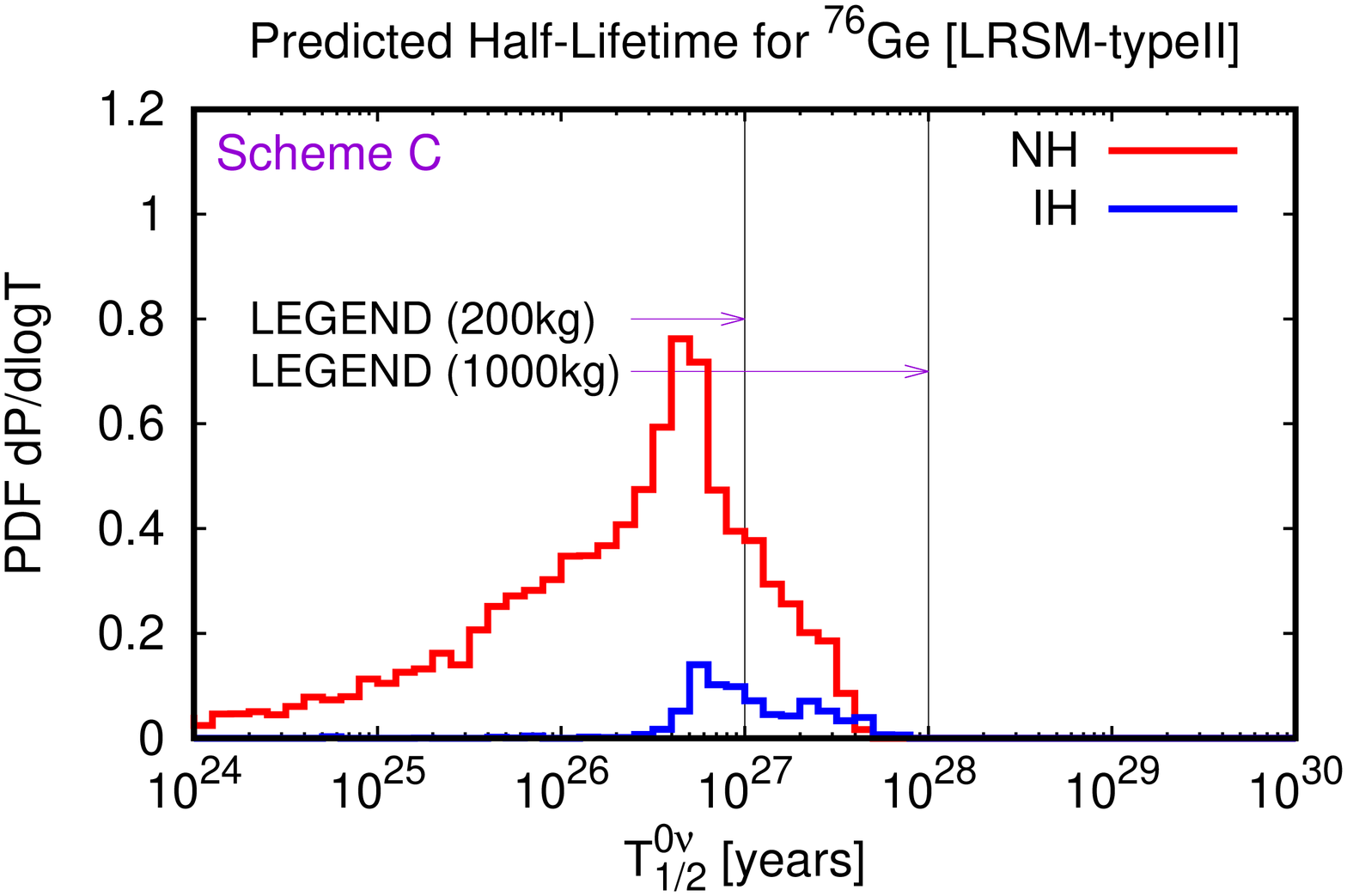}
\includegraphics[width=4cm,height=0.32\textwidth,angle=-90]{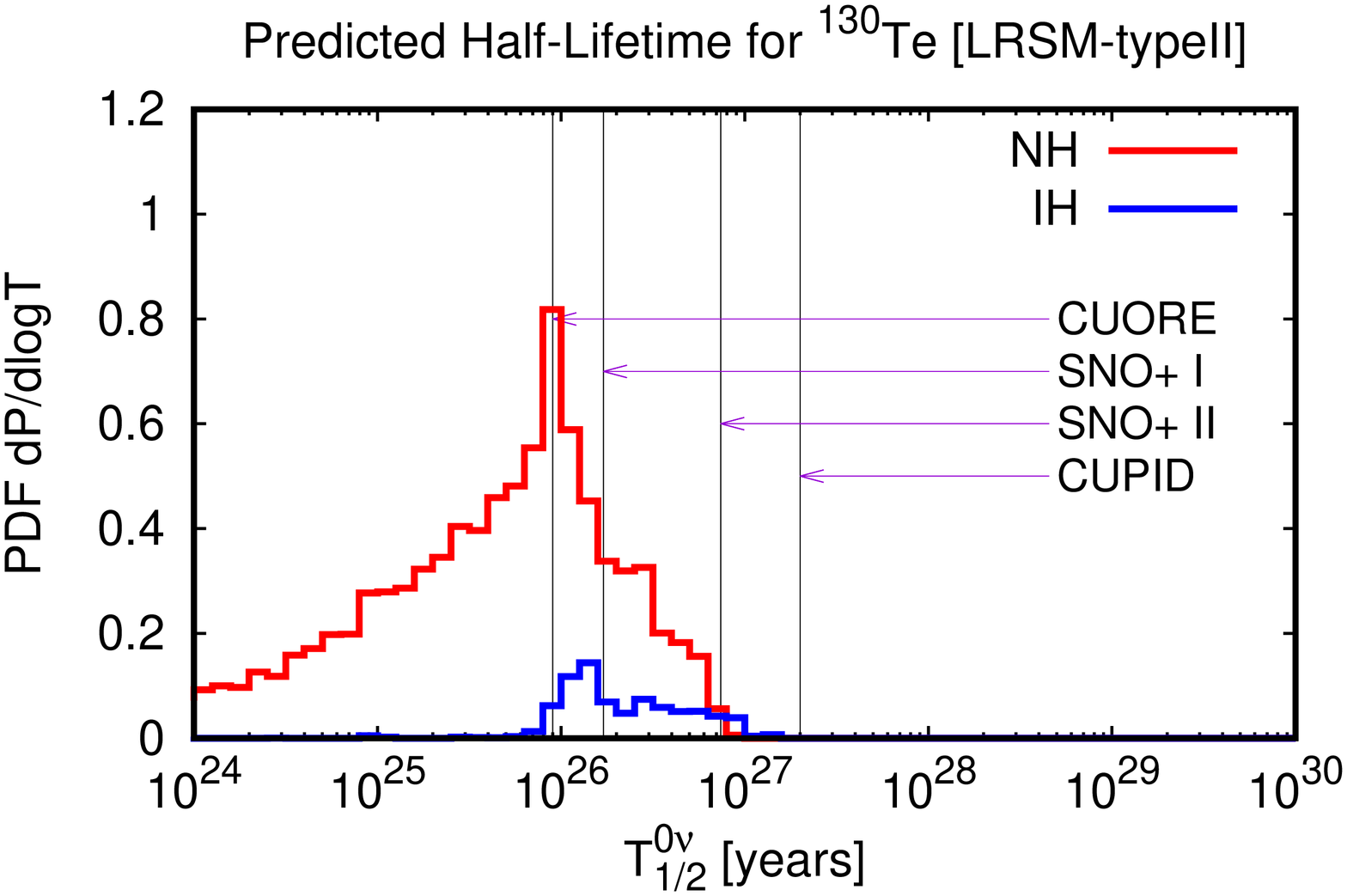}
\includegraphics[width=4cm,height=0.32\textwidth,angle=-90]{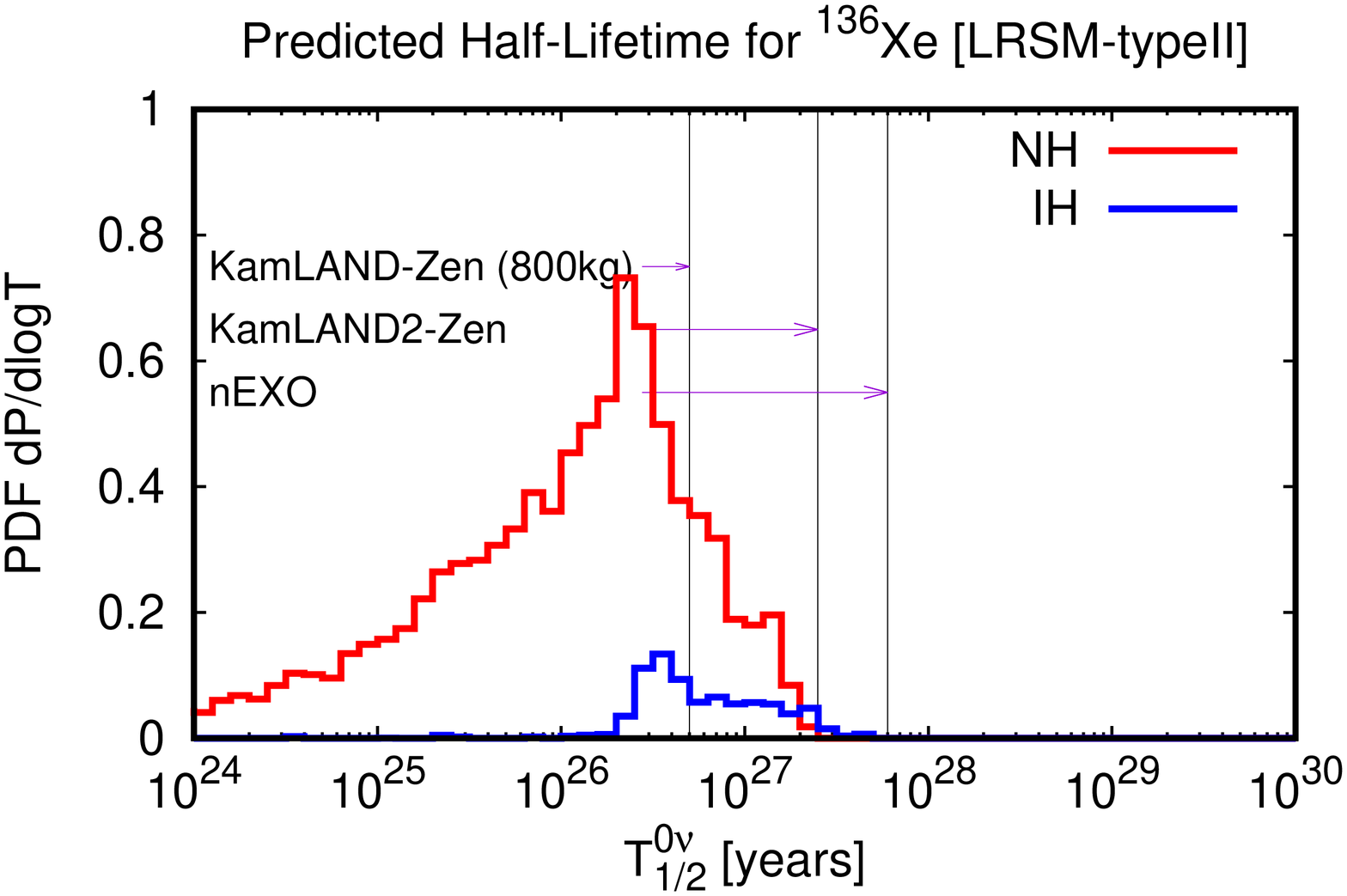}
\includegraphics[width=4cm,height=0.32\textwidth,angle=-90]{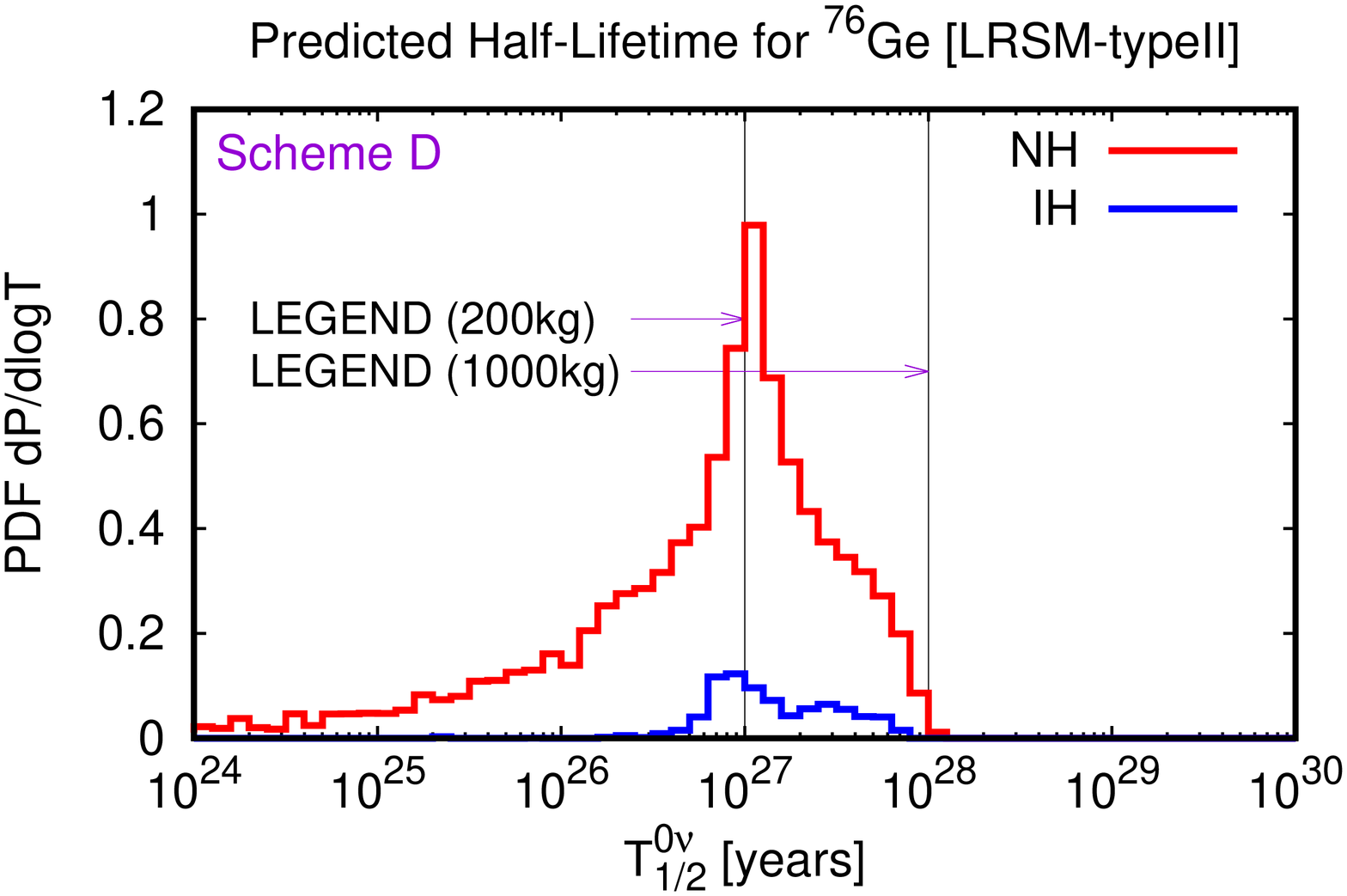}
\includegraphics[width=4cm,height=0.32\textwidth,angle=-90]{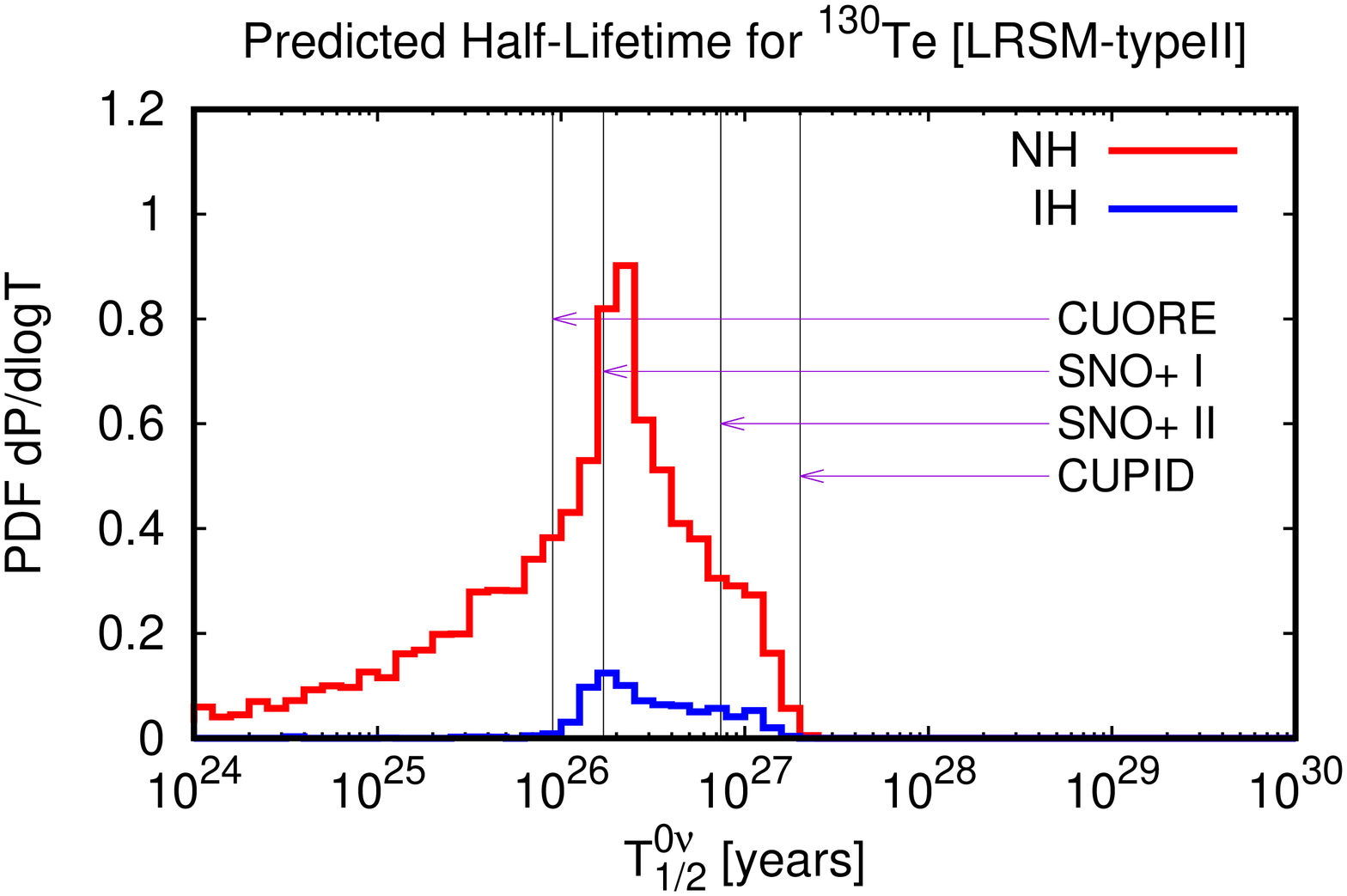}
\includegraphics[width=4cm,height=0.32\textwidth,angle=-90]{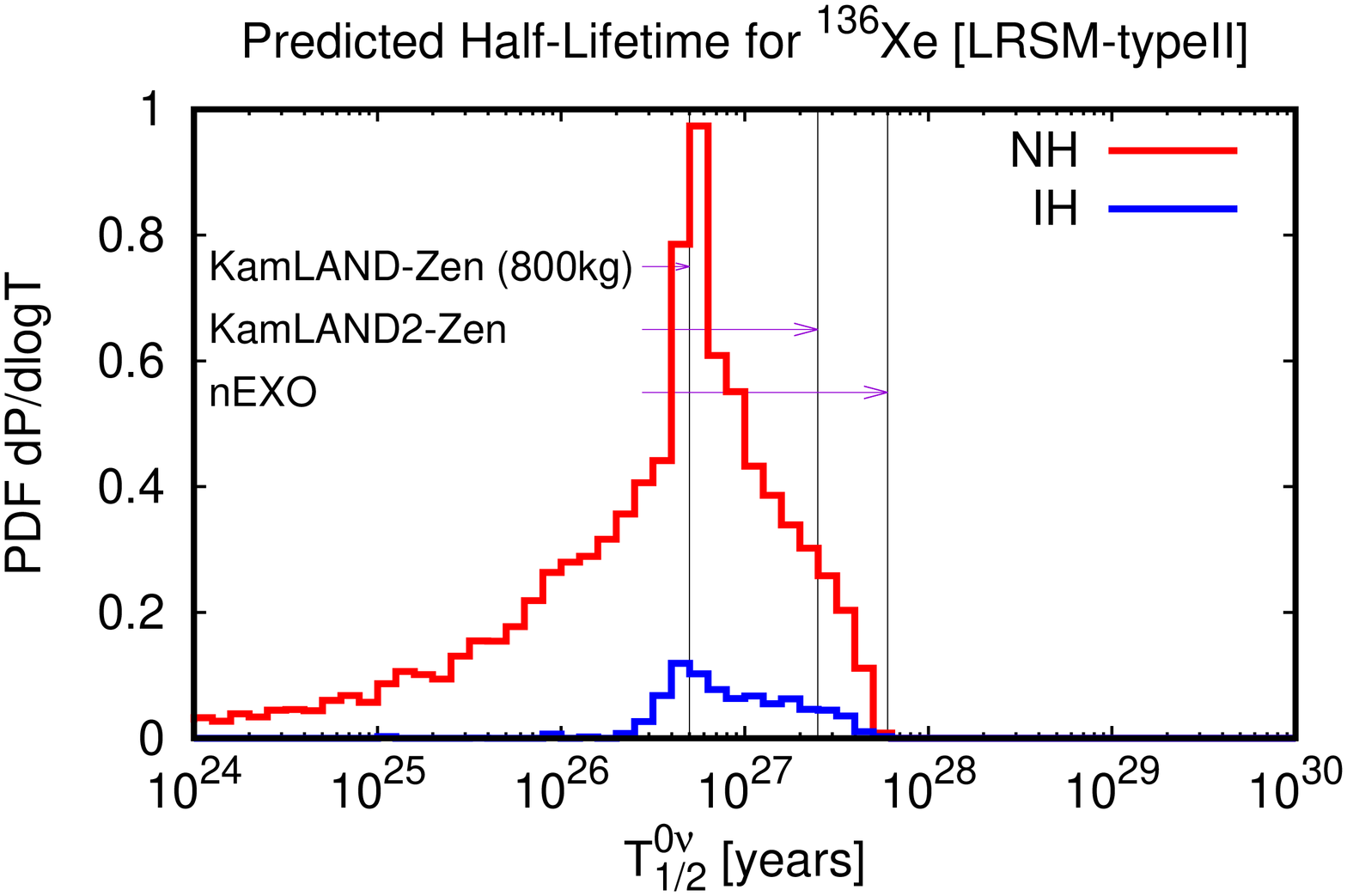}
\caption{The predicted distribution of neutrinoless double beta decay half-lives in left-right symmetric theories 
with type II seesaw dominance. Four different scenarios are plotted, Scheme A ($g_R/g_L = \frac 23$, $M_{W_R} = 2.5$ TeV, , $M_{N_{\rm heavy}} = 1$ TeV), Scheme B ($g_R/g_L = \frac 23$, $M_{W_R} = 3.5$ TeV, , $M_{N_{\rm heavy}} = 1$ TeV), Scheme C ($g_R/g_L = 1$, $M_{W_R} = 2.5$ TeV, $M_{N_{\rm heavy}} = 1$ TeV) and Scheme D ($g_R/g_L = 1$, $M_{W_R} = 2.5$ TeV, $M_{N_{\rm heavy}} = 2$ TeV).}
\label{fig:sampleThalf-LRSM}
\end{figure}
Here $\mee^{\rm light}$ is the standard 3-neutrino effective mass from Section \ref{sec:nu}, $g_L$ ($g_R$) is the gauge 
coupling of the left-handed (right-handed) gauge sectors, and $M_{W_L}$ $(M_{W_R})$ are the masses of the 
left-handed (right-handed) gauge bosons. We will first assume for illustration $g_R/g_L = \frac 23$, $M_{W_R} = 2.5$ TeV and a 
heaviest heavy neutrino mass $M_{\rm heavy}$ of 1 TeV\footnote{Note that for heavy neutrino masses below 100 MeV the expression (\ref{eq:lr}) does not apply. This would correspond to a ratio of smallest to largest mass of about $10^{-4}$, which according to \gfig{fig:massratio} is quite unlikely and has no visible effect on the plots to follow.}. 
\gfig{fig:mee-LRSM} shows the predicted distribution of $\mee^{\rm LR}$, as well as the individual 
terms, using oscillation and cosmology data. For comparison with the light neutrino
contribution, the ratio of NMEs  is not included in the
heavy neutrino contribution (red)
in \gfig{fig:mee-LRSM}, but only included in the combined $\mee^{\rm LR}$ for
$\Ge$, $\Te$, and $\Xe$. 
While $\mee^{\rm light} \propto U_{ei}^2 \, m_i $ for NH is bounded from above, 
the new term proportional to $|U_{ei}^2 /m_i|$ 
is bounded from below. The situation is the opposite for IH. Thus, there is again a different 
behavior with respect to the standard case \cite{Tello:2010am}. 
The combined result $\mee^{\rm LR}$ is then bounded 
from below for both NH and IH, consequently also a lower limit on neutrino mass arises in such scenarios \cite{Dev:2013vxa}. 
Note that $\meeLRSM$ is dominated by heavy (light) neutrinos for NH (IH), respectively.  
Correspondingly, the amplitude can not be arbitrarily small  and half-lives 
longer than $\sim 10^{28}$ yrs are unlikely, as shown in the upper panel of 
\gfig{fig:sampleThalf-LRSM}. In that figure we also show half-life expectations for 
other values of the gauge parameters $g_R$ and $M_{W_R}$. Scheme A is the example 
described so far ($g_R/g_L = \frac 23$, $M_{W_R} = 2.5$ TeV, $M_{\rm heavy} = 1$ TeV), while we further define 
Scheme B ($g_R/g_L = \frac 23$, $M_{W_R} = 3.5$ TeV, $M_{\rm heavy} = 1$ TeV), Scheme C ($g_R/g_L = 1$, $M_{W_R} = 2.5$ TeV, $M_{\rm heavy} = 1$ TeV) and Scheme D ($g_R/g_L = 1$, $M_{W_R} = 2.5$ TeV, , $M_{\rm heavy} = 2$ TeV). The dependence on the heaviest heavy neutrino mass is weaker than the dependence on $g_R$ or $M_{W_R}$, see Eq.\ \eqref{eq:lr}.

Comparing NH and IH, the half-life  for TeV-scale type II dominated left-right symmetry  
tends to have a longer tail in the lower end for NH, which reflects the larger 
effective mass $\langle m_{ee}\rangle^{\rm LR}$ for this case. 
The remaining parameter region has large overlap between NH and IH. 
For this particular example, a half-life limit of $1\times 10^{27}$ years as achievable by LEGEND-200 would cover 21.1\% of the expected normal ordering range and 33.9\% of the inverted ordering one. CUPID, with a sensitivity of $2\times10^{27}$ years, could cover 88.1\% of the normal and 99.96\% of the inverted ordering. The nEXO experiment, with a possible limit of $6\times 10^{27}$ years, could over 90.0\% of the expected normal ordering range and 99.8\% of the inverted ordering one. 
The coverage of the different experimental projects for the two mass orderings is summarized in \gtab{tab:res}. 
Obviously those numbers depend on the values of the right-handed neutrinos and gauge boson. 
Choosing for instance Scheme B gives 6.8\% and 31.8\% for LEGEND-200, as well as 57.5\% and 99.7\% for nEXO (recall that for IH the standard contribution dominates). Finally, the necessary half-lives to cover 95\% of the expected half-lives are summarized in \gtab{tab:res1}.


\begin{table}[t]
\centering
\begin{tabular}{c|c|ccc}
 Isotope & Experiment  & standard & LR & sterile \\
\hline
                     \multirow{2}{*}{${}^{76}{\rm Ge}$}     & LEGEND 200 ($1 \times 10^{27}$ y) & 3.77\% (39.2\%)   & 21.1\% (33.9\%)  & 7.63\% (45.4\%) \\
                                    & LEGEND 1000 ($1 \times 10^{28}$ y)  & 38.3\% (98.0\%)   & 87.7\% (100\%)   & 96.7\% (91.5\%) \\
\hline
 \multirow{4}{*}{${}^{130}{\rm Te}$} & CUORE ($9 \times 10^{25}$ y)    & 0.65\% (6.21\%)   & 12.9\% (0.91\%)  & 0.69\% (13.1\%)\\
                                     & SNO$+$ I ($1.7 \times 10^{26}$ y)    & 2.31\% (20.8\%)   & 19.2\% (14.9\%)  & 4.13\% (30.0\%) \\
                                     & SNO$+$ II ($7.4 \times 10^{26}$ y)    & 15.3\% (74.6\%)   & 64.9\% (79.1\%)  & 51.4\% (73.2\%) \\
                                     & CUPID ($2 \times 10^{27}$ y)    & 32.7\% (93.5\%)   & 88.1\% (99.96\%) & 88.4\% (88.2\%) \\
\hline
 \multirow{3}{*}{${}^{136}{\rm Xe}$} 
                                     & KamLAND-Zen 800 ($5 \times 10^{26}$ y)  & 3.04\% (25.0\%)   & 20.8\% (19.3\%)  & 6.94\% (31.1\%)\\
                                     & KamLAND2-Zen ($2.5 \times 10^{27}$ y)  & 18.1\% (77.3\%)   & 71.8\% (84.0\%)  & 57.6\% (75.2\%) \\
                                     & nEXO ($6 \times 10^{27}$ y)   & 33.6\% (92.5\%)   & 90.0\% (99.8\%)  & 86.4\% (87.7\%)
\end{tabular}
\caption{The relative exclusion probability of the projected experimental sensitivities for normal (inverted) mass ordering and the different scenarios under consideration: the standard approach of 3 Majorana neutrinos, the addition of eV-scale 
sterile neutrinos, and TeV-scale left-right symmetry (with $g_R/g_L = \frac 23$, $M_{W_R} = 2.5$ TeV and a 
heaviest heavy neutrino mass of 1 TeV). 
Experimental limits correspond to $3\sigma$ half-life sensitivity after 4 years of running, taken mostly from \cite{fut}. 
The NMEs from Tables \ref{tab:NME} and \ref{tab:NME-LRSM} were used. }
\label{tab:res}
\end{table}

\section{\label{sec:concl}Summary}
In this paper we have obtained half-life expectations for neutrinoless double beta decay in three scenarios that all have different dependence on low energy neutrino parameters: the standard approach with exchange of three active neutrinos, the case when an eV-scale sterile neutrino is added, and TeV-scale left-right symmetric theories with type II seesaw dominance. 
Available probability distributions of oscillation parameters, mass constraints from cosmology and nuclear matrix element calculations were applied.  
The discovery potential of upcoming experiments using $^{76}$Ge, $^{130}$Te and $^{136}$Xe can be estimated from the obtained half-life distributions, and the outcome is summarized in \gtab{tab:res}. 
The percentage of values that can be reached for the normal or inverted mass ordering is quite different for the different cases, and illustrates both the subtlety of extracting definite physics from double beta decay, as well as the rich physics potential of upcoming experiments. 

\begin{table}[t]
\centering
\begin{tabular}{c|ccc}
95\% exclusion $\log_{10}(T_{1/2})$ for NH (IH) & standard & LR & sterile \\
\hline
${}^{76}{\rm Ge}$  & 29.8 (27.8) & 28.2 (27.8) & 27.9 (28.2) \\
${}^{130}{\rm Te}$ & 29.2 (27.4) & 27.5 (27.1) & 27.5 (27.7) \\
${}^{136}{\rm Xe}$ & 29.7 (27.9) & 27.9 (27.6) & 28.0 (28.2)
\end{tabular}
\caption{The necessary experimental sensitivities to fully exclude the normal (inverted) mass ordering for the different scenarios under consideration: the standard approach of 3 Majorana neutrinos, the addition of eV-scale 
sterile neutrinos, and TeV-scale left-right symmetry (with $g_R/g_L = \frac 23$, $M_{W_R} = 2.5$ TeV and a 
heaviest heavy neutrino mass of 1 TeV). 
Experimental limits correspond to $3\sigma$ half-life sensitivity after 4 years of running, taken mostly from \cite{fut}. 
The NMEs from Tables \ref{tab:NME} and \ref{tab:NME-LRSM} were used. }
\label{tab:res1}
\end{table}

\section*{Acknowledgments}

WR is supported by the DFG with grant RO 2516/6-1 in the Heisenberg program.

\bibliographystyle{apsrev4-1}
\bibliography{ref}

\end{document}